\documentclass[aps,prb,reprint,superscriptaddress,longbibliography,floatfix,raggedbottom]{revtex4-2}
\usepackage{amsmath,amssymb,bm,graphicx,xcolor,booktabs,microtype}
\usepackage[colorlinks=true,citecolor=blue,linkcolor=blue,urlcolor=blue]{hyperref}
\makeatletter
\AtBeginDocument{\immediate\write\@auxout{\string\citation{apsrev42Control}}}
\makeatother
\graphicspath{{figures/}}
\newcommand{\Lio}{\mathcal{L}}
\newcommand{\Diss}{\mathcal{D}}
\newcommand{\Tr}{\operatorname{Tr}}
\newcommand{\GapF}{\Gamma_{\mathrm{filt}}}
\newcommand{\rhoss}{\rho_{\mathrm{ss}}}

\begin{document}

\title{\texorpdfstring{Anomalously fast heating of ground states\\
in dephased $SU(N)$ spin systems}{Anomalously fast heating of ground states in dephased SU(N) spin systems}}

\author{Tairan Lou}
\affiliation{School of Physics, Nankai University, Tianjin 300071, China}
\author{Junxiao Chen}
\affiliation{School of Physics, Nankai University, Tianjin 300071, China}
\author{Xiang-Ping Jiang}
\affiliation{School of Physics, Hangzhou Normal University, Hangzhou, Zhejiang 311121, China}
\author{Mingdi Xu}
\email{2120230195@mail.nankai.edu.cn}
\affiliation{School of Physics, Nankai University, Tianjin 300071, China}
\author{Lei Pan}
\email{panlei@nankai.edu.cn}
\affiliation{School of Physics, Nankai University, Tianjin 300071, China}

\date{October 18, 2026}

\begin{abstract}
We identify a strong quantum Mpemba effect in which a many-body ground state heats faster than every Gibbs state at finite positive temperature under the same dissipative dynamics within a fixed-color sector.  In ferromagnetic $SU(N)$ exchange systems with local dephasing, exact coherence shells confine the ground-state evolution and exclude all slower Liouvillian modes.  Its relaxation rate is fixed by the bath, independently of system size, exchange graph, and integrability.  We derive an energy-resolved sign criterion that guarantees a slower thermal contribution throughout the positive-temperature range.  An exact finite-chain solution proves this rate inversion, and symmetry-resolved calculations confirm the criterion in larger spin-one chains, including nonintegrable exchange.  The zero-temperature endpoint of a continuous Gibbs family can therefore be dynamically isolated by its exact exclusion from slow relaxation channels.
\end{abstract}

\maketitle

\section{Introduction}

A cold quantum state can heat toward an infinite-temperature stationary state faster than an initially warmer state.  This reversal of thermal ordering is a form of the quantum Mpemba effect (QME).  Classical studies established a statistical description of anomalous relaxation~\cite{Mpemba1969,Jeng2006,LuRaz2017,Klich2019,KumarNature2020,KumarPNAS2022,TezaReview2026}.  Quantum coherence, correlations, and symmetry introduce further possibilities~\cite{AresReview2025,YuSymmetry2025,Moroder2024,VuHayakawa2025,Strachan2025}.  In the strong quantum Mpemba effect (SQME), the initially farther state has a larger asymptotic decay rate~\cite{Carollo2021,Kochsiek2022,NavaEgger2024,Chatterjee2024,CaldasPires2026,SalibaDrumond2025,LonghiGiant2026}.  Many-body examples include symmetry restoration in integrable chains~\cite{AresAsymmetry2023,MurcianoXY2024,Chalas2024,Rylands2024}, random circuits~\cite{LiuRandom2024,TurkeshiCircuits2025,KlobasTranslation2025,AresCircuits2025}, spatially inhomogeneous systems~\cite{LiuLocalized2025,LonghiRare2026}, long-range models~\cite{YamashikaLongRange2026,HallamLongRange2026}, and chaotic dynamics without global symmetries~\cite{BhoreNoSymmetry2025}.  Experiments have observed both symmetry-restoration QME and dissipative strong Mpemba relaxation~\cite{Joshi2024,Shapira2024,ZhangStrong2025,XuProcessor2026,ZhanCoherence2026}.  Open-system mechanisms include interactions~\cite{GrafJPCM2025,WangWeiNJP2026,AliBoseHubbard2026}, external noise~\cite{ZhaoNoise2026}, and environmental memory~\cite{ZhangNonMarkovEP2026,StrachanImpurity2026}.

Comparing a ground state with finite-temperature states of the same Hamiltonian gives a precise setting for this problem.  All preparations belong to one Gibbs family and evolve under the same heating channel.  Dissipative spin chains exhibit such acceleration near criticality~\cite{WeiCritical2026}, with long-range exchange~\cite{SongLongRange2026}, and through exact restrictions on the accessible decay modes~\cite{WeiComb2026,XuRydberg2026}.  The observed rate depends on the slowest Liouvillian mode excited by a preparation, so excluding slow modes from the ground state can make it relax rapidly.  A comparison with the entire thermal family, however, also requires control of the thermal overlaps.  A slow amplitude can vanish at an isolated temperature, which a finite temperature grid may miss.  Symmetry alone need not resolve this issue: it excludes nontrivial representations but can leave slow scalar modes, while fast symmetry restoration does not by itself establish fast relaxation of the full state~\cite{AresReview2025,Caceffo2024}.  This distinction is especially important near zero temperature.  Thermal density matrices approach the ground state continuously, yet their asymptotic rates may remain different if the slow amplitude vanishes only at the endpoint.  Establishing this behavior requires both an exact ground-state trajectory and a condition on the full positive-temperature family.

Here we study ferromagnetic $SU(3)$ exchange chains subject to local color dephasing.  In a fixed-color sector, the ground state is a color-Dicke state: its configuration probabilities already equal those of the final mixed state, while its coherences account for the initial departure from stationarity.  Exchange symmetry organizes these coherences into exact Liouvillian shells and prevents them from generating a population imbalance.  The ground-state rate is consequently fixed by dephasing.  The protection follows from compatibility between exchange and color dephasing, although the full generator is not $SU(3)$ invariant.  We then derive a sign criterion, expressed through complete energy-band traces of a slow left eigenoperator, that makes a slower thermal amplitude nonzero at every positive temperature.  An exact finite-chain construction proves this rate inversion, and larger-chain calculations verify the criterion beyond a temperature scan.  The same shell algebra holds on arbitrary exchange graphs and extends to $SU(N)$, where odd coherence shells occur for three or more occupied colors.  Direct energy and trace-distance evolution, symmetry-resolved level statistics, and weak-anisotropy calculations determine how the acceleration appears in finite systems and how it is affected by imperfections.

The paper is organized as follows.  Section~\ref{sec:model} defines the model and thermal preparations.  Section~\ref{sec:colorcomb} derives the exact ground-state dynamics, and Sec.~\ref{sec:thermal} establishes the positive-temperature criterion.  Section~\ref{sec:numerics} presents numerical results, followed by experimental considerations in Sec.~\ref{sec:experiment}.  Section~\ref{sec:outlook} concludes with an outlook.

\section{Model and heating protocol}
\label{sec:model}

We consider a periodic ferromagnetic $SU(3)$ exchange chain,
\begin{equation}
 H=-J_1\sum_jP_{j,j+1}-J_2\sum_jP_{j,j+2},
 \label{eq:su3j1j2}
\end{equation}
where $P_{ij}$ exchanges the three local states, $J_1>0$, and $J_2\ge0$.  We use $J_1$ as the energy unit.  The Hamiltonian is invariant under global $SU(3)$ rotations and conserves each color occupation.  For a spin-one realization,
\begin{equation}
 P_{ij}=\bm S_i\cdot\bm S_j+(\bm S_i\cdot\bm S_j)^2-I.
 \label{eq:spin1-permutation}
\end{equation}
Using Eq.~\eqref{eq:spin1-permutation}, Eq.~\eqref{eq:su3j1j2} has equal ferromagnetic bilinear and biquadratic couplings, up to a constant~\cite{Laeuchli2006,Dai2022}.  Nearest-neighbor permutation exchange is Bethe integrable~\cite{Uimin1970,Lai1974,Sutherland1975}.  The next-nearest-neighbor term preserves the exchange symmetry and the ground state.  We use symmetry-resolved level statistics to select a nonintegrable coupling, allowing the relaxation to be compared without changing the conserved color occupations.

The construction extends to $N$ internal states on a connected graph, with $H=-\sum_{i<j}J_{ij}P_{ij}$ and $J_{ij}\ge0$.  Local color dephasing drives the system according to the Lindblad equation~\cite{Lindblad1976,Gorini1976,BreuerBook,Diehl2008,Muller2012,Harrington2022,Fazio2025}
\begin{equation}
 \begin{aligned}
 \partial_t\rho=\Lio(\rho)&=-i[H,\rho]+\Diss(\rho),\\
 \Diss(\rho)&=\gamma\sum_{j,a}\left(P_{j,a}\rho P_{j,a}
                  -\tfrac12\{P_{j,a},\rho\}\right).
 \end{aligned}
 \label{eq:lindblad}
\end{equation}
Here $P_{j,a}=|a\rangle_j\langle a|$, and the colors are $m_z=1,0,-1$ for spin one.  Equal projector rates damp every unequal-color single-site coherence at $\gamma$.  The bath selects this basis and does not preserve the full continuous $SU(3)$ symmetry of $H$.  We set $\hbar=k_B=1$.  The total occupations $\bm N=(N_1,\ldots,N_N)$ remain conserved.  Within a fixed-color sector, the stationary state is unique and equals $\rhoss=I_{\bm N}/\mathcal D_{\bm N}$, where $\mathcal D_{\bm N}=L!/\prod_aN_a!$.  We compare states only within the same sector, so they all approach this common infinite-temperature state.

\begin{figure}[!tbp]
\includegraphics[width=\columnwidth]{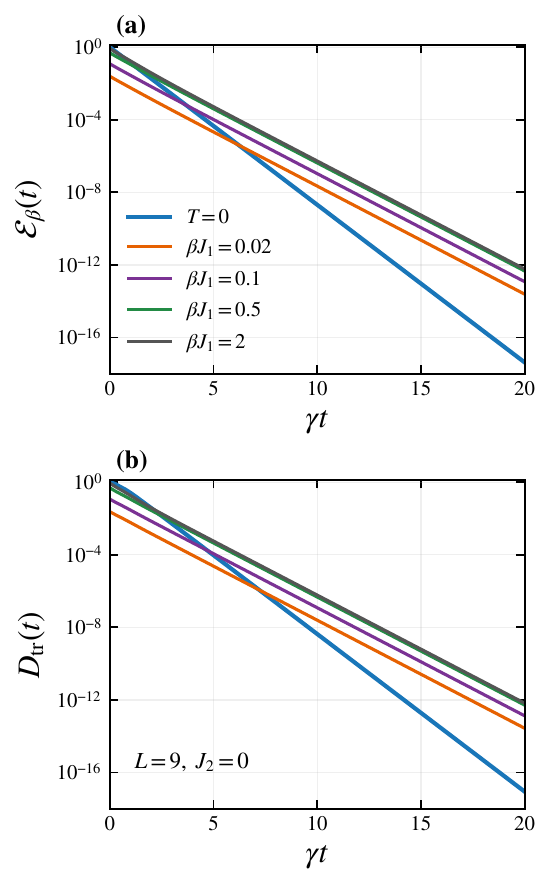}
\caption{\label{fig:heating}Anomalously fast ground-state heating at $L=9$, $\bm N=(3,3,3)$, $J_2=0$, and $\gamma/J_1=0.4$.  (a) Commonly normalized energy deviation $\mathcal E_\beta$ [Eq.~\eqref{eq:heating-diagnostics}], computed by propagating the Gibbs family in the exact symmetry quotient.  (b) Trace distance of the evolved states, evaluated in all translation blocks.  Thermal trajectories are propagated directly through $\gamma t=20$; ground-state curves use the exact shell solution.  The finite temperatures are $\beta J_1=0.02,0.1,0.5,2$.  The zero-temperature state starts farther away in both diagnostics and overtakes each plotted thermal state; its energy follows $\mathcal E_D=e^{-2\gamma t}$.  The continuous-temperature statement follows from Sec.~\ref{sec:thermal}.}
\end{figure}

The initial Gibbs family and its zero-temperature limit are
\begin{equation}
 \begin{aligned}
 \rho_\beta&=e^{-\beta H}/Z_\beta,\qquad Z_\beta=\Tr e^{-\beta H},\\
 \rho_D&=\lim_{\beta\to\infty}\rho_\beta=|D_{\bm N}^{(N)}\rangle\langle D_{\bm N}^{(N)}|,\\
 |D_{\bm N}^{(N)}\rangle&=\frac{\sum_{x\in\Omega_{\bm N}}|x\rangle}
 {\sqrt{\mathcal D_{\bm N}}}.
 \end{aligned}
 \label{eq:color-dicke}
\end{equation}
Every exchange leaves the Dicke state invariant.  It therefore minimizes each ferromagnetic bond and reaches $E_0=-\sum_{i<j}J_{ij}$.  Connected bonds generate all site permutations, whose common symmetric state is unique at fixed color occupations.  The Dicke state is thus the ground state of the Hamiltonian used to define $\rho_\beta$.  A finite positive-temperature Gibbs state is full rank and has a smaller trace distance from $\rhoss$ than this pure ground state.  Its energy lies strictly between $E_0$ and $E_\infty=\Tr(H)/\mathcal D_{\bm N}$.  We monitor both the trace distance and a normalized energy deviation,
\begin{equation}
 \begin{aligned}
 D_{\rm tr}(t)&=\tfrac12\|\rho(t)-\rhoss\|_1,\\
 \mathcal E_\beta(t)&=\frac{|\Tr[H(\rho_\beta(t)-\rhoss)]|}{E_\infty-E_0}.
 \end{aligned}
 \label{eq:heating-diagnostics}
\end{equation}
We use the same energy denominator in Eq.~\eqref{eq:heating-diagnostics} for every preparation.  The evolving state generally leaves the Gibbs family, so temperature labels its preparation only.  Figure~\ref{fig:heating}(a) shows the inversion in energy, and Fig.~\ref{fig:heating}(b) shows it in trace distance.  The ground state starts farther from stationarity in both measures and subsequently overtakes the plotted thermal states.  We first derive its exact trajectory, then establish when a slower thermal contribution survives throughout the positive-temperature range.

\section{Exact relaxation of the zero-temperature ground state}
\label{sec:colorcomb}

The origin of the fast ground-state relaxation becomes apparent in the basis selected by the bath.  A state $|x\rangle=|a_1a_2\cdots a_L\rangle$ specifies the color at each site, with the prescribed total occupations $\bm N$.  Equation~\eqref{eq:color-dicke} assigns the same amplitude to every allowed configuration.  The ground-state density matrix and the final stationary state are therefore
\begin{equation}
 \begin{aligned}
 \rho_D&=\frac1{\mathcal D_{\bm N}}\sum_{x,y\in\Omega_{\bm N}}|x\rangle\langle y|,\\
 \rhoss&=\frac1{\mathcal D_{\bm N}}\sum_{x\in\Omega_{\bm N}}|x\rangle\langle x|,\\
 \langle x|\rho_D|x\rangle&=\langle x|\rhoss|x\rangle=\frac1{\mathcal D_{\bm N}}.
 \end{aligned}
 \label{eq:stationary-populations}
\end{equation}
The common diagonal in Eq.~\eqref{eq:stationary-populations} gives the probability of every many-body configuration.  This is a stronger equality than uniform one-site densities or fixed total color numbers.  It also concerns a different basis from the energy eigenbasis, in which the zero-temperature state occupies only the ground-energy level.  The off-diagonal elements distinguish the two states: for $x\ne y$, $\langle x|\rho_D|y\rangle=1/\mathcal D_{\bm N}$, while $\rhoss$ has no coherence.  The pure ground state has initial trace distance $1-1/\mathcal D_{\bm N}$, the largest possible value in this sector.  Equal configuration probabilities thus coexist with a large departure from the stationary quantum state.  We refer to these configuration probabilities as populations below.

Dephasing directly removes the off-diagonal coherences.  Whether the initial uniform populations remain unchanged depends on the Hamiltonian, which can generally convert coherence into a population imbalance.  In the present model, the evolution of the ground state closes within a set of uniform coherence sums, preventing this conversion.  To construct them, note that
\begin{equation}
 \Diss^\dagger(|x\rangle\langle y|)
       =-\gamma d_H(x,y)|x\rangle\langle y|,
 \label{eq:dephasing}
\end{equation}
where $d_H(x,y)$ is the number of sites at which the two configurations differ.  The bath damps a coherence at Hamming distance $m$ at the rate $m\gamma$.  Summing all matrix units at a given distance defines
\begin{equation}
 \begin{aligned}
 O_m&=\sum_{\substack{x,y\in\Omega_{\bm N}\\d_H(x,y)=m}}|x\rangle\langle y|,\\
 \Lio(O_m)&=\Lio^\dagger(O_m)=-m\gamma O_m.
 \end{aligned}
 \label{eq:suncomb}
\end{equation}
The second line follows by simultaneously exchanging two sites in $x$ and $y$.  Their distance is unchanged, and $P_{ij}$ permutes the terms in $O_m$, giving $P_{ij}O_mP_{ij}=O_m$.  Each uniform shell therefore commutes with every exchange bond, while the dissipator acts on it with a single rate.  Individual matrix units at the same distance need not commute with $H$.  The shell identity holds for arbitrary real exchange couplings; positive couplings and connectedness ensure the ground-state interpretation and a unique stationary state.  Since the shell span is invariant under both $\Lio$ and $\Lio^\dagger$, it is a reducing subspace even when the full generator is non-normal~\cite{Prosen2008,Minganti2018}.

The Dicke density matrix is the uniform sum of these shells.  Its exact evolution is
\begin{equation}
 \rho_D(t)=\frac{1}{\mathcal D_{\bm N}}
 \left[I_{\bm N}+\sum_{m\ge2}e^{-m\gamma t}O_m\right].
 \label{eq:exactdynamics}
\end{equation}
All shells with $m>0$ are off diagonal, so $\langle x|\rho_D(t)|x\rangle=1/\mathcal D_{\bm N}$ at every time.  The Hamiltonian commutator vanishes along this trajectory: dephasing removes coherence without generating a population imbalance.  Two different configurations in the same fixed-color sector cannot differ at only one site.  If at least two colors are occupied, $O_2\ne0$, and the leading decay exponent of any norm-induced state distance is consequently $2\gamma$.  Higher shells contribute at earlier times, so the trace distance need not follow a single exponential throughout its evolution.

The energy nevertheless obeys a single exponential at all times.  The diagonal part of $H$ has a constant expectation along Eq.~\eqref{eq:exactdynamics}, while every off-diagonal exchange matrix element connects two configurations at distance two.  Only $O_2$ therefore contributes to the energy deviation,
\begin{equation}
 \begin{aligned}
 E_D(t)-E_\infty&=(E_0-E_\infty)e^{-2\gamma t},\\
 \Gamma_D&=2\gamma.
 \end{aligned}
 \label{eq:ground-energy-law}
\end{equation}
The system heats as dephasing removes the coherences that lower its exchange energy, even though the configuration probabilities do not change.  This gives a measurable ground-state heating law, with a rate independent of size, exchange graph, and integrability.  Appendix~\ref{app:proof} establishes the norm-independent asymptotic exponent.  With two occupied colors, mismatches occur in pairs and only even shells are allowed.  For three or more occupied colors, cyclic rearrangements add odd shells beginning at $m=3$, as seen in the shell weights of Fig.~\ref{fig:comb}(a), calculated in Appendix~\ref{app:counting}.  Figure~\ref{fig:comb}(b) compares ground-state trace distances at different sizes and shows the common leading exponent.  These full-state Liouvillian eigenoperators describe a different quantity from subsystem entanglement asymmetry in non-Abelian symmetry-restoration QME~\cite{FujimuraNonAbelian2026}.

\begin{figure}[!tbp]
\includegraphics[width=\columnwidth]{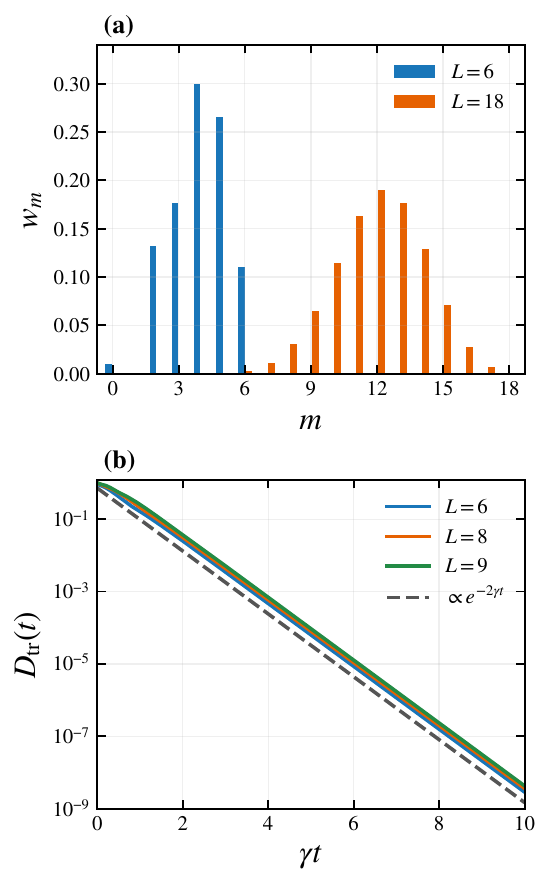}
\caption{\label{fig:comb}Exact coherence-shell weights and ground-state relaxation.  (a) Squared shell coefficients $w_m=n_m/\mathcal D_{\bm N}^2$ at balanced $L=6$ and $18$, obtained by exact integer counting in Eq.~\eqref{eq:shell-counting}.  The large-size weights follow from shell counting and do not require the Liouvillian spectrum.  Both even and odd teeth occur.  (b) Dicke-state trace distance for $L=6,8,9$, with nearest-balanced occupations.  The dashed line indicates the exact leading exponent $2\gamma$.}
\end{figure}

\begin{figure*}[t]
\includegraphics[width=\textwidth]{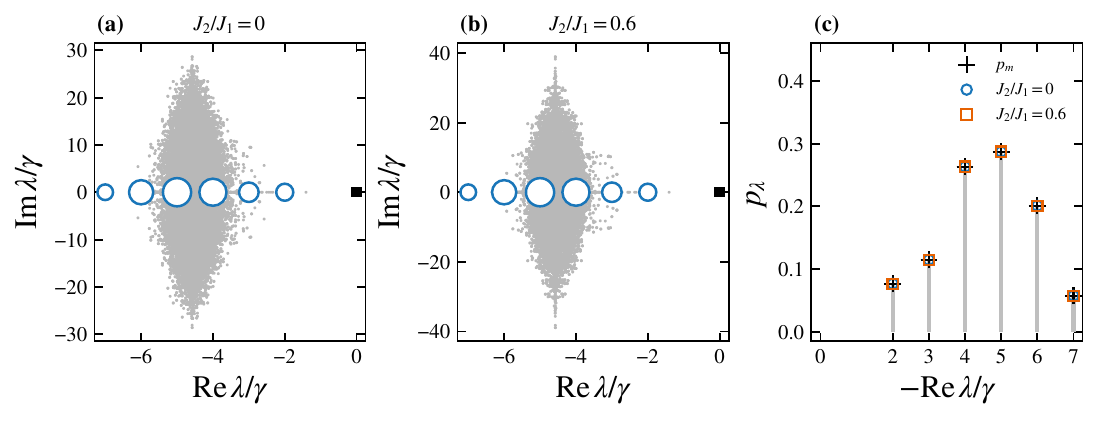}
\caption{\label{fig:modeprojections}Dicke-state support in the complete Liouvillian spectrum at $L=7$, $\bm N=(3,2,2)$, and $\gamma/J_1=0.4$.  (a,b) All $\mathcal D_{\bm N}^2$ eigenvalues for $J_2/J_1=0$ and $0.6$ are obtained by diagonalizing every translation and cyclic-color character block and combining their complete spectra.  Gray points show the surrounding spectrum; blue circles mark the occupied comb teeth, with area proportional to $p_\lambda$.  The black square is the stationary eigenvalue, removed from $\rho_D-\rhoss$.  (c) Projection weights evaluated by polynomial spectral projectors on the verified reducing shell span agree with the exact weights.  Appendix~\ref{app:counting} describes the projector construction and an independent five-site dual-basis calculation.  The arbitrary-size support follows from Eq.~\eqref{eq:suncomb}.}
\end{figure*}

The spectral projection shows explicitly which decay modes are absent.  For $X_D=\rho_D-\rhoss$, let $\Pi_\lambda$ project onto the full generalized eigenspace at $\lambda$.  The normalized weights are
\begin{equation}
 \begin{aligned}
 p_\lambda&=\|\Pi_\lambda X_D\|_2^2/\|X_D\|_2^2,\\
 p_{-m\gamma}&=\frac{n_m}{\mathcal D_{\bm N}(\mathcal D_{\bm N}-1)},\quad m\ge2,
 \end{aligned}
 \label{eq:dicke-projection}
\end{equation}
and vanish off the comb.  Distinct shells are Hilbert--Schmidt orthogonal, so these weights sum to unity.  This probability interpretation follows from the reducing shell subspace; it is not a general property of projections for a non-normal generator.  Figures~\ref{fig:modeprojections}(a,b) compare the complete spectra for two exchange choices.  Although the surrounding eigenvalues change, the Dicke state has support on the same real comb.  Figure~\ref{fig:modeprojections}(c) resolves its projection weights and compares them with the shell formula.  Degenerate modes are projected together, making the result independent of the eigenvector basis.  Appendix~\ref{app:counting} derives the normalized mode amplitudes and their explicit $SU(3)$ counting formula, and gives the spectral projectors.  Appendix~\ref{app:proof} establishes the logarithmic decay exponent for a general finite-dimensional generator.

\section{Thermal relaxation at every finite positive temperature}
\label{sec:thermal}

To compare the exact ground-state rate with the entire thermal family, we need a slower mode whose amplitude never vanishes at a positive temperature.  Excited energy levels allow a Gibbs state to explore modes outside the ground-state shell space, but their contributions to a given slow mode can cancel.  We address this possibility using complete energy-band traces.  Choose a Hermitian left eigenoperator $Q$ with $\Lio^\dagger Q=-\Gamma_sQ$ and $0<\Gamma_s<2\gamma$.  Its expectation decays at $\Gamma_s$, so a nonzero thermal expectation certifies a slow contribution.  Stationarity of $\rhoss$ gives $\Tr Q=0$, and the exact ground-state shells give $\Tr(Q\rho_D)=0$.  Resolve $H$ into distinct energies $E_0<E_1<\cdots<E_M$ with projectors $P_k$, and define
\begin{equation}
 a_k=\Tr(QP_k),\qquad A_k=\sum_{j=1}^k a_j.
 \label{eq:cumulative-coefficients}
\end{equation}
Each $a_k$ in Eq.~\eqref{eq:cumulative-coefficients} includes the entire degenerate eigenspace at $E_k$.  Ground-state exclusion gives $a_0=0$, and tracelessness gives $A_M=0$.  The coefficients $a_k$ can have either sign.  A sufficient condition preventing a Boltzmann-weighted cancellation is that all interior cumulative sums satisfy $A_k\ge0$, with at least one strict inequality.  The overall sign of $Q$ can be reversed.  Finite summation by parts then yields
\begin{equation}
 \begin{aligned}
 F_Q(\beta)&\equiv\Tr(Q\rho_\beta)=\frac{\sum_{k=1}^{M}a_ke^{-\beta E_k}}{Z_\beta}\\
 &=\frac{1}{Z_\beta}\sum_{k=1}^{M-1}A_k
       \left(e^{-\beta E_k}-e^{-\beta E_{k+1}}\right)>0.
 \end{aligned}
 \label{eq:abel-positive}
\end{equation}
Every Boltzmann difference is positive for $\beta>0$, proving a nonzero slow amplitude throughout the continuous temperature range.  Writing $\Gamma_{\rm filt}$ for the slowest rate visible to the initial state, we obtain
\begin{equation}
 \Gamma_{\rm filt}(\rho_\beta)\le\Gamma_s<2\gamma=\Gamma_D,
 \qquad 0<\beta<\infty.
 \label{eq:all-temperature-rate}
\end{equation}
The rate comparison also follows directly from a bound on the trace distance.  The expectation $\Tr[Q\rho_\beta(t)]=F_Q(\beta)e^{-\Gamma_s t}$ gives
\begin{equation}
 D_{\rm tr}[\rho_\beta(t),\rhoss]\ge
 \frac{F_Q(\beta)}{2\|Q\|_\infty}e^{-\Gamma_s t}.
 \label{eq:thermal-distance-bound}
\end{equation}
The ground-state distance decays at $2\gamma$, so their distance ratio tends to zero for each fixed finite positive temperature.  Together with the initial ordering, this establishes the strong effect.  The slow mode necessarily has a population component: the Hamiltonian part contributes no real Hilbert--Schmidt dissipation, and every off-diagonal matrix unit in the fixed-color sector is damped at least at $2\gamma$.  The general spectral definition of $\Gamma_{\rm filt}$ is given in Appendix~\ref{app:selection}.  Appendix~\ref{app:alltemperature} quantifies the population content and gives the observable-response condition for the same mode to appear in energy.

\begin{figure*}[t]
\includegraphics[width=\textwidth]{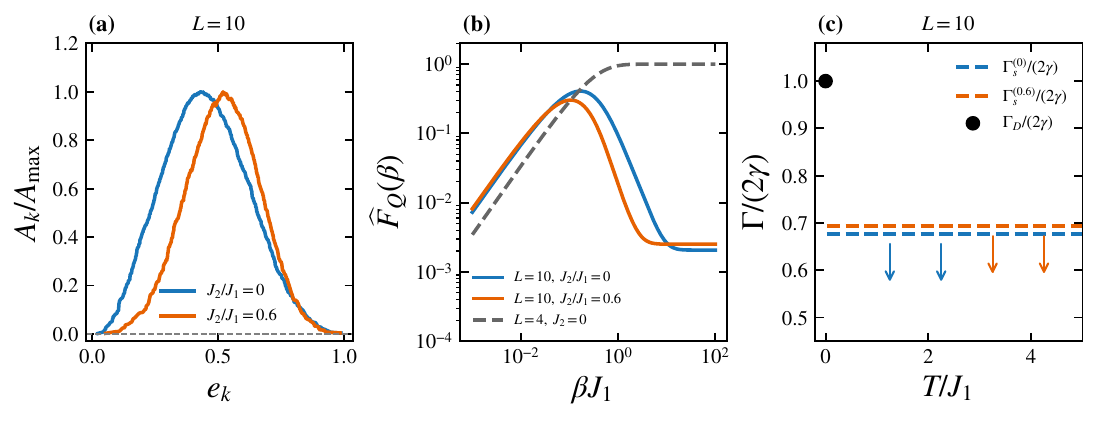}
\caption{\label{fig:alltemperature}Continuous-temperature criterion at $L=10$.  (a) Energy-band cumulative weights for both exchange choices, normalized by $A_{\rm max}$, with $e_k=(E_k-E_0)/(E_M-E_0)$.  Every interior cumulative is positive.  (b) Rescaled overlap $\widehat F_Q=Z_\beta e^{\beta E_1}F_Q/A_{\rm max}$ from the same complete bands.  The four-site expression is retained as a dashed analytic reference.  (c) Temperature-independent upper bounds $\Gamma_{\rm filt}(\rho_\beta)\le\Gamma_s^{(J_2)}$ obtained from the two ten-site witnesses, compared with the exact ground-state rate.  Downward arrows indicate the bound; the lines are not fitted thermal exponents.  The four-site algebraic proof is given in Appendix~\ref{app:alltemperature}.  The point at $T=0$ represents the Dicke preparation.  All panels use $\gamma/J_1=0.4$.}
\end{figure*}

For an exact realization (Appendix~\ref{app:thermal-exact}), consider the nearest-neighbor four-site chain in the $(2,1,1)$ sector at $\gamma/J_1=2/5$.  An algebraic construction of the slow left mode gives, after a fixed normalization,
\begin{equation}
 F_Q(\beta)=\frac{e^{2\beta J_1}
 (1-e^{-2\beta J_1})(1+R e^{-2\beta J_1})}{Z_\beta}.
 \label{eq:exact-all-temperature-overlap}
\end{equation}
Here $R>0$, $\Gamma_s<2\gamma$, and the energy response is nonzero.  The expression is positive for every $\beta>0$, proving both state-distance and energy-rate inversion for the whole thermal family.  Exact root counting establishes that this is the leading nonstationary mode in the common symmetry sector.  Its rate is thus shared by all positive-temperature preparations and changes at the zero-temperature endpoint.  In larger chains (Appendix~\ref{app:thermal-numerics}), the complete energy-band coefficients satisfy the cumulative sign condition for both exchange choices [Fig.~\ref{fig:alltemperature}(a)].  Their Boltzmann-weighted sums remain positive throughout the temperature range [Fig.~\ref{fig:alltemperature}(b)], and the corresponding rates bound the thermal decay below the ground-state rate [Fig.~\ref{fig:alltemperature}(c)].  These finite-size sign checks extend the continuous-temperature comparison through the verified sizes; an arbitrary-size sign theorem remains open.  The exact construction and numerical details are given in Appendix~\ref{app:alltemperature}.

The thermal condition is essential even though the ground-state shells hold on every exchange graph.  On a uniform complete graph, the thermal density matrix is also invariant under all site permutations and evolves within the dephasing shells.  Both preparations then have leading rate $2\gamma$, so there is no ground-state rate advantage.  This example, derived in Appendix~\ref{app:thermal-complete-graph}, separates the exact ground-state solution from the additional spectral and overlap conditions needed for an SQME.

\section{Many-body relaxation and observation times}
\label{sec:numerics}

\begin{figure*}[t]
\includegraphics[width=\textwidth]{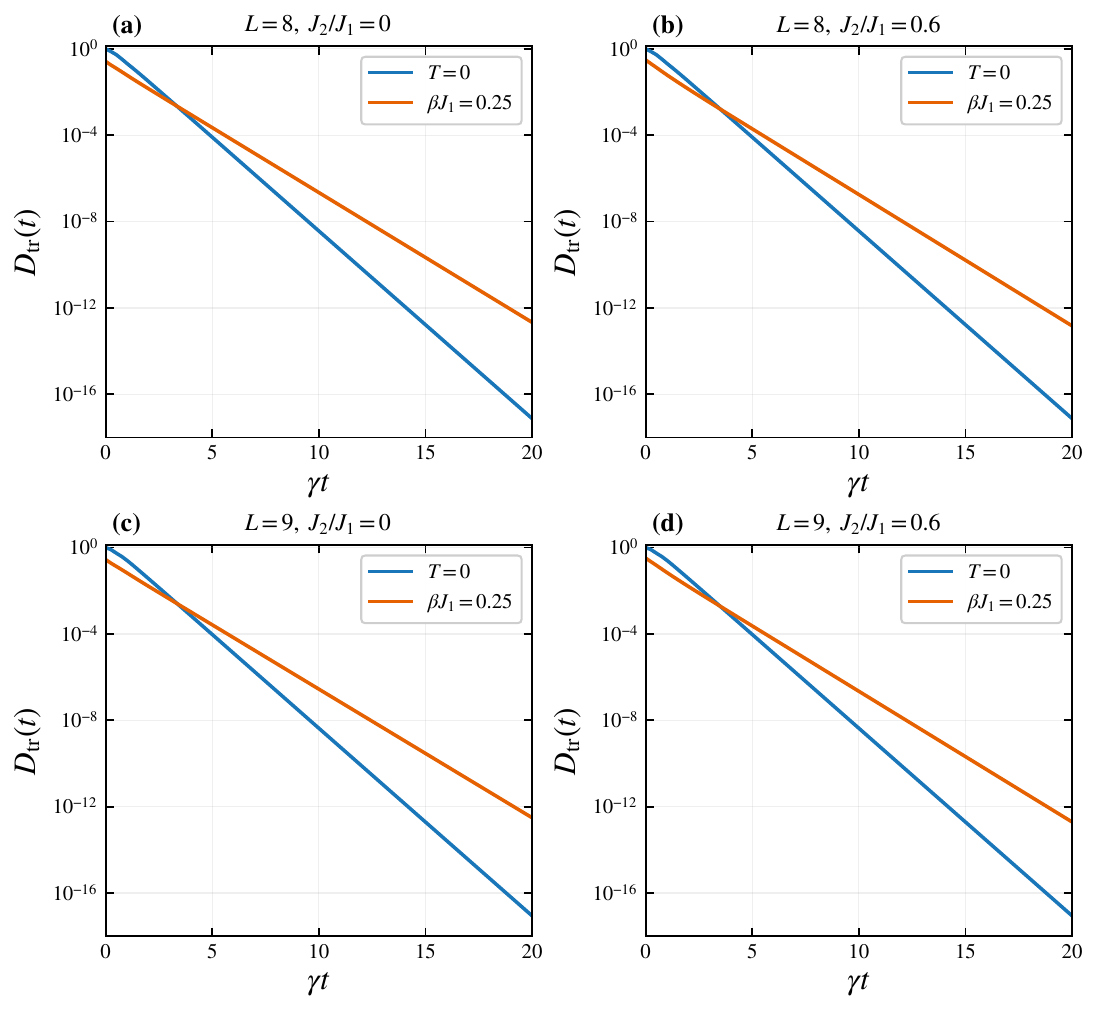}
\caption{\label{fig:dynamics}Trace-distance evolution from ground-state and thermal preparations.  Rows correspond to $L=8$ and $9$, with nearest-balanced occupations; columns use $J_2/J_1=0$ and $0.6$.  The Dicke state starts farther from stationarity and overtakes the Gibbs state in each case.  All panels use $\gamma/J_1=0.4$ and $\beta J_1=0.25$.  Gibbs deviations are propagated directly through $\gamma t=20$, while the ground-state curves follow Eq.~\eqref{eq:exactdynamics}.  The independent complete energy-band criterion extends the positive-temperature comparison through $L=10$.}
\end{figure*}

\begin{figure*}[t]
\includegraphics[width=\textwidth]{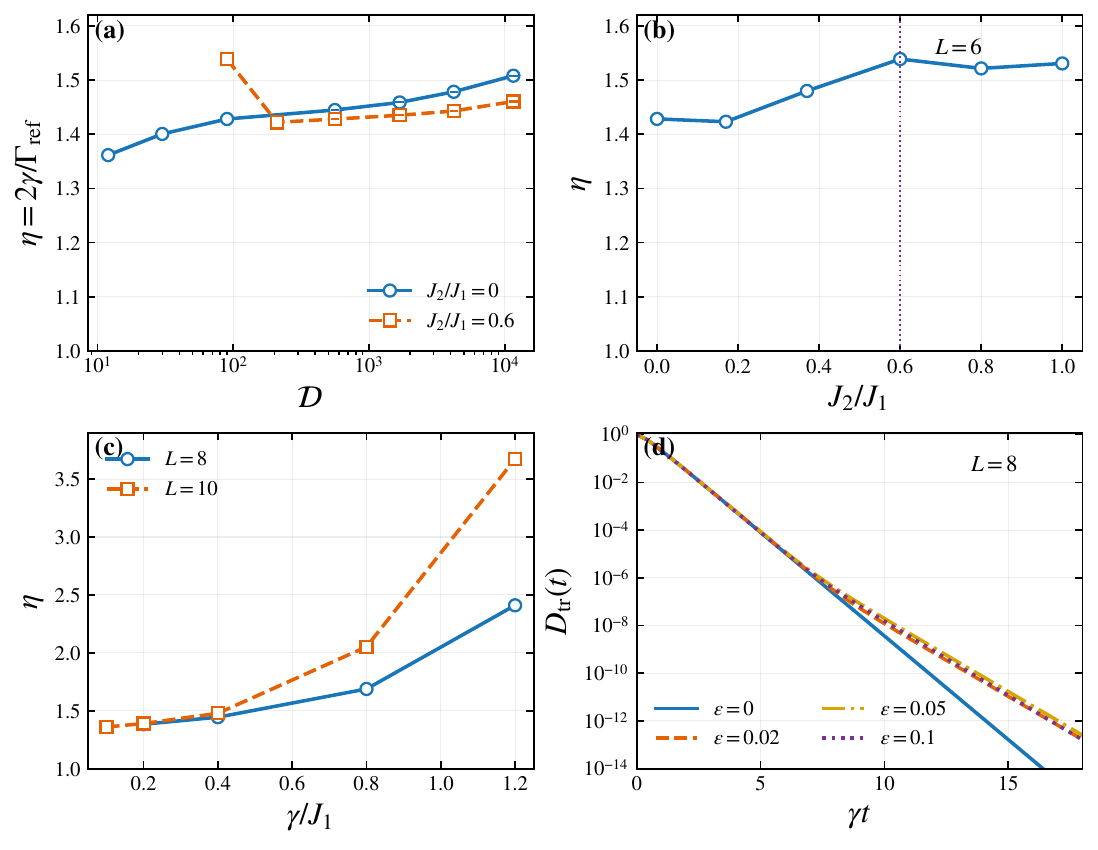}
\caption{\label{fig:scaling}Size, coupling, and dephasing dependence of the rate inversion.  (a) Rate ratio $\eta=2\gamma/\Gamma_{\rm ref}$ through $L=11$, plotted against the fixed-color Hilbert-space dimension.  Smaller sizes use exact Gibbs evolution; the larger sequence uses high-temperature Gibbs-tangent Hilbert--Schmidt tails.  (b) Exchange scan at $L=6$.  (c) Dephasing scans at $L=8$ and $10$.  (d) Trace-distance trajectories at $L=8$ showing the slow tail restored by the number-conserving perturbation $V_\epsilon=\epsilon\sum_jS_j^zS_{j+1}^z$ added to Eq.~\eqref{eq:su3j1j2}.  Fit-window sensitivities and preparation definitions are given in Appendix~\ref{app:numerics}.}
\end{figure*}

Direct evolution tests how the rate separation appears in state distances.  Figures~\ref{fig:dynamics}(a,b) show the Gibbs and ground-state trajectories at $L=8$ for the two exchange choices; Figs.~\ref{fig:dynamics}(c,d) show the corresponding $L=9$ results.  The ground state starts farther from stationarity and overtakes the thermal state, with the late-time slope fixed by $2\gamma$.  To extend the size comparison, we propagate the linear high-temperature Gibbs direction with a matrix-free Krylov method and extract its Hilbert--Schmidt tail [Fig.~\ref{fig:scaling}(a)].  At overlapping sizes, its fitted rates agree closely with the finite-temperature trace-distance rates.  This calculation extends the finite-size relaxation data; the statement for all positive temperatures comes from the independent energy-band criterion.  Preparation definitions, fitting windows, and convergence checks are given in Appendix~\ref{app:numerics}.

The ground-state rate stays fixed as the exchange couplings change, while the thermal rate varies [Fig.~\ref{fig:scaling}(b)].  Increasing the dephasing strength further separates the rates in the calculated range [Fig.~\ref{fig:scaling}(c)].  Figure~\ref{fig:scaling}(d) shows the slow tail restored by weak exchange anisotropy; its observation interval is analyzed below.

\begin{figure*}[t]
\includegraphics[width=\textwidth]{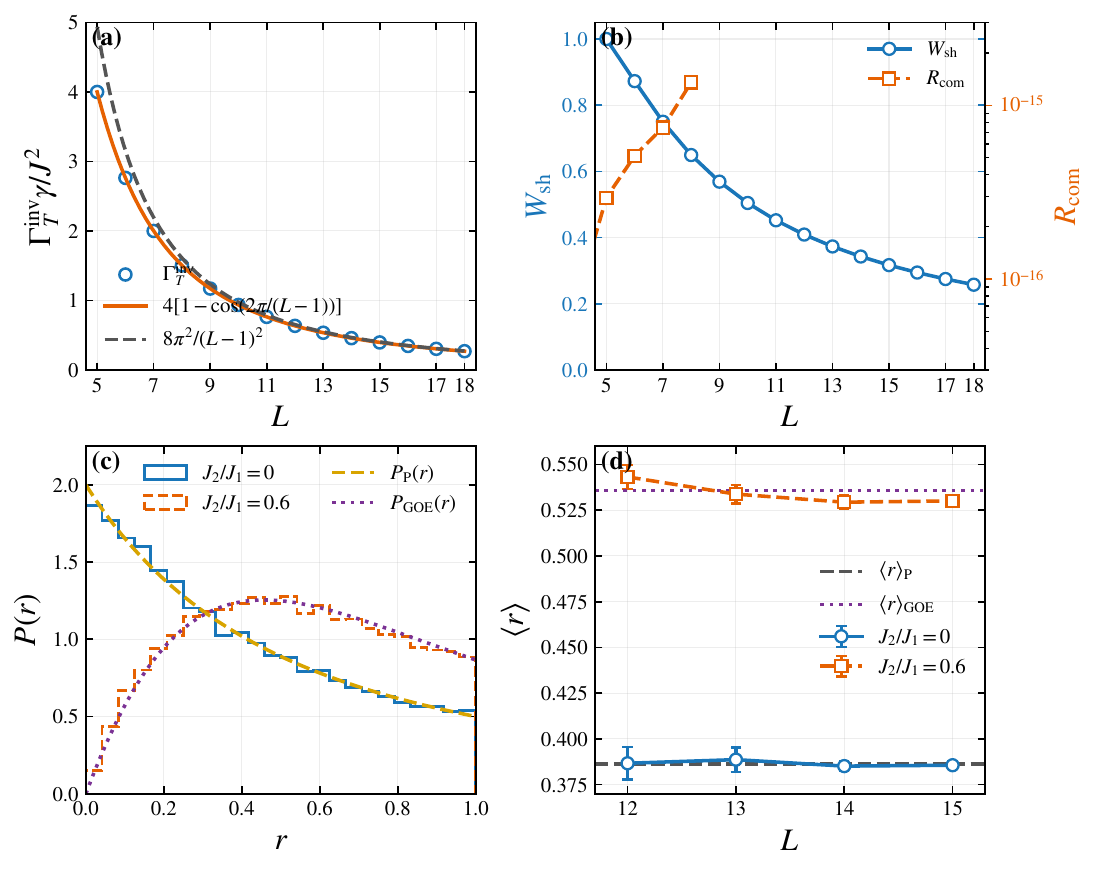}
\caption{\label{fig:thermochaos}Strong-dephasing relaxation and symmetry-resolved Hamiltonian statistics.  (a) Fully invariant Zeno population rate for $L=5$--$18$ and the branch in Eq.~\eqref{eq:giantscaling}.  (b) High-temperature population-mode weight on the same sequence, with a shell-commutator check on random exchange graphs.  (c) Adjacent-gap-ratio distribution at $L=15$ after resolving the $SU(3)$ representation and momentum.  (d) Block-averaged ratio for $L=12$--$15$.  Panels (a,b) refer to the effective strong-dephasing generator of Appendix~\ref{app:zenogap}.  Panels (c,d) characterize the Hamiltonians used for the finite-$\gamma$ relaxation comparison.}
\end{figure*}

A complementary description of the nearest-neighbor chain is available when $\gamma\gg J_1$, the quantum Zeno limit.  Rapidly decaying coherences can then be eliminated, leaving an effective classical exchange generator for the populations~\cite{Znidaric2010,Medvedyeva2016}.  Its relaxation scale is $J_1^2/\gamma$, whereas the exact ground-state rate remains proportional to $\gamma$.  Appendix~\ref{app:zenogap} derives this limiting generator and its slow symmetric branch.  Figures~\ref{fig:thermochaos}(a,b) show the branch and its high-temperature weight, which decreases with size.  The limit gives an additional explanation of slow population relaxation; neither the ground-state solution nor the finite-$\gamma$ thermal criterion assumes it.  The decrease of the weight also means that the limiting rate alone does not determine a thermodynamic crossing time.

To check the role of integrability, we resolve a fixed $SU(3)$ irreducible representation and individual momentum blocks before evaluating Hamiltonian level statistics (Appendix~\ref{app:levelstats}).  Nearest-neighbor exchange gives Poisson statistics, while the selected next-nearest-neighbor coupling gives a GOE-like distribution [Figs.~\ref{fig:thermochaos}(c,d)].  Reflection combined with complex conjugation preserves a nonzero-momentum block and squares to unity, fixing the random-matrix comparison.  The rate inversion therefore persists when the Hamiltonian has nonintegrable spectral statistics, with the same exact ground-state shell structure.

\begin{figure*}[t]
\includegraphics[width=\textwidth]{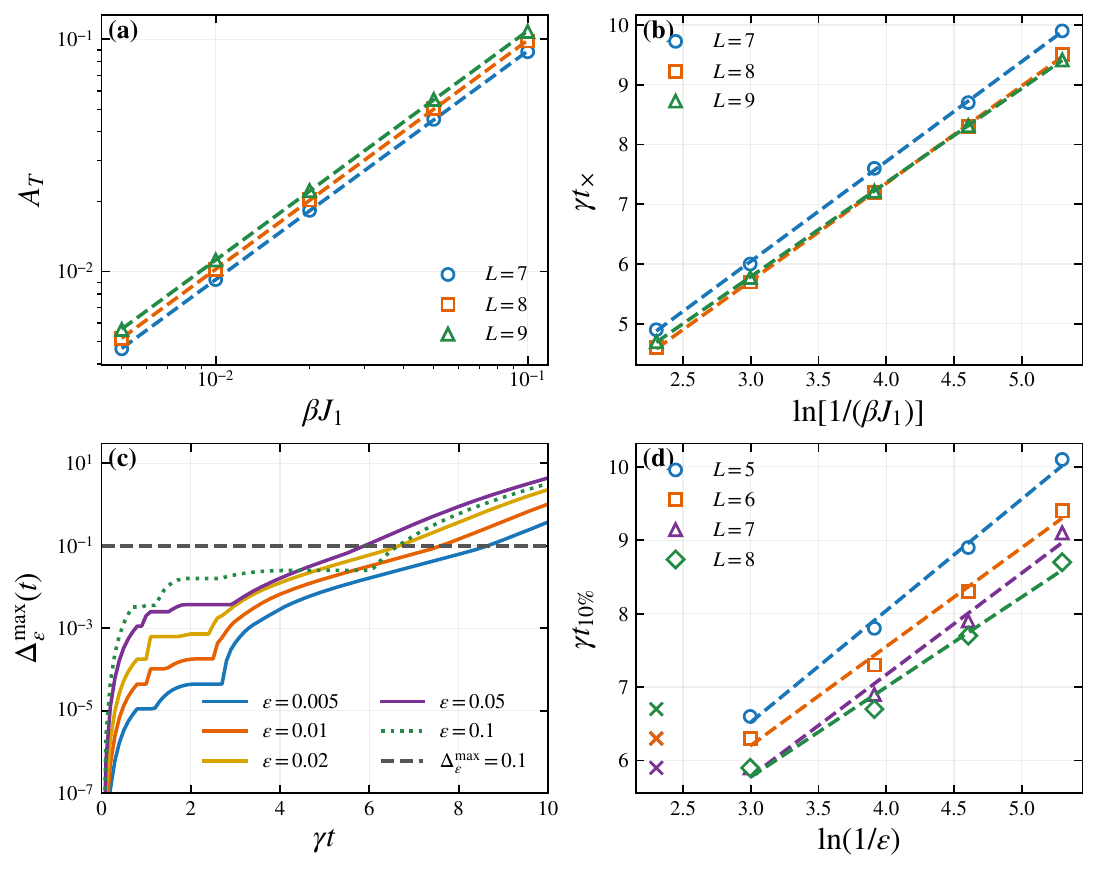}
\caption{\label{fig:windows}Temperature dependence and the acceleration interval under weak anisotropy.  (a,b) Thermal slow-tail amplitude and crossing delay for $L=7,8,9$.  (c) Relative deviation from protected Dicke dynamics at $L=8$ under the specified weak anisotropy.  (d) Acceleration windows for $L=5$--$8$.  The horizontal line in (c) marks the $10\%$ relative-deviation threshold.  Dashed fits in (a,b,d) are described in Appendix~\ref{app:numerics}.}
\end{figure*}

The amplitude of a slow contribution determines when the asymptotic advantage becomes observable.  At high temperature, $\rho_\beta-\rhoss=-\beta(H-\langle H\rangle_0 I)/\mathcal D_{\bm N}+O(\beta^2)$, so the leading slow amplitude vanishes linearly as $\beta\to0$ [Fig.~\ref{fig:windows}(a)].  This produces the logarithmic crossing delay in Fig.~\ref{fig:windows}(b).  Near zero temperature, the slow amplitude is exponentially small in $\beta$ when its first contributing excitation is gapped.  Its late-time dominance can then be difficult to resolve, despite a nonzero amplitude at every fixed positive temperature.  The onset of that tail and the first crossing between two nearby preparations are distinct times.  Likewise, the asymptotic rate ordering does not order every fixed-accuracy stopping time, since a hot reference may already be within the chosen tolerance initially.  The sensitivity to mixedness also appears in other QME measures~\cite{AresMixed2025,Caceffo2024}.

Weak exchange anisotropy introduces a further time scale.  The perturbation $V_\epsilon=\epsilon\sum_jZ_jZ_{j+1}$, with $Z=\operatorname{diag}(1,0,-1)$, breaks the shell commutation and restores a slow amplitude $a_s(\epsilon)$.  Comparing this term with the fast component gives
\begin{equation}
 t_\epsilon\simeq\frac{\ln|A_f/a_s(\epsilon)|}{\Gamma_f-\Gamma_s}.
 \label{eq:imperfection-window}
\end{equation}
Figure~\ref{fig:windows}(c) tracks the departure from protected dynamics.  If $a_s$ vanishes as a power of $\epsilon$, Eq.~\eqref{eq:imperfection-window} gives a logarithmically growing acceleration interval as the perturbation is reduced, as shown in Fig.~\ref{fig:windows}(d).  Its width depends on the induced amplitude; a universal first-order dependence on $\epsilon$ is not required.  The exact infinite-time separation belongs to the unperturbed exchange model, while the perturbed trajectories exhibit an extended finite interval of faster heating.

\subsection{Experimental considerations}
\label{sec:experiment}

The exchange energy provides a direct observable for the heating comparison.  In a spin-one realization, its bilinear and biquadratic contributions can be inferred from color-resolved two-site correlations.  The exact ground-state law in Eq.~\eqref{eq:ground-energy-law} then supplies a reference without fitting a late-time slope.  A digital qutrit array can encode the three colors and implement exchange through two-qutrit gates generated by $P_{ij}$.  Universal qudit control and entangling gates have been demonstrated in trapped-ion processors~\cite{Ringbauer2022,Hrmo2023}; controlled dissipation is also available in atomic platforms~\cite{Barontini2013,Patil2015,Bouganne2020}.  A number-conserving symmetric circuit prepares the Dicke state, while a purified or randomized charge-conserving preparation can supply a thermal reference.  Energy and color occupations give independent preparation checks.  Existing ion, superconducting, and photonic QME experiments~\cite{Joshi2024,ZhangStrong2025,XuProcessor2026,ZhanCoherence2026} demonstrate relevant relaxation measurements.  The present protocol additionally requires exchange dynamics and local color dephasing, which can be generated by independent diagonal phase kicks after each exchange layer,
\begin{equation}
 \begin{aligned}
 U_\phi&=\exp\!\left[-i\sum_{j,a}\phi_{j,a}P_{j,a}\right],\\
 \overline{\phi_{j,a}\phi_{k,b}}&=\gamma\delta t\,\delta_{jk}\delta_{ab}.
 \end{aligned}
 \label{eq:random-phases}
\end{equation}
Averaging Eq.~\eqref{eq:random-phases} reproduces Eq.~\eqref{eq:lindblad} to leading order in the Trotter step, and local Ramsey measurements calibrate the dephasing rate.  The phases must vary independently between sites: spatially uniform relative phases give collective noise.

An observable comparison also requires sensitivity to the leading modes of both preparations~\cite{BaguiObservable2025}.  Besides energy, a common readout can combine the distance-two coherence shell $C_2=O_2$ with $E_{\rm cl}=\sum_{j,a}P_{j,a}P_{j+1,a}$.  The embedded two-color quadratures satisfy
\begin{equation}
 |a_i b_j\rangle\langle b_i a_j|+\mathrm{H.c.}
 =\tfrac12(X_i^{ab}X_j^{ab}+Y_i^{ab}Y_j^{ab}),
 \label{eq:exchange-quadratures}
\end{equation}
where $X^{ab}=|a\rangle\langle b|+|b\rangle\langle a|$ and $Y^{ab}=-i(|a\rangle\langle b|-|b\rangle\langle a|)$.  Using Eq.~\eqref{eq:exchange-quadratures}, six quadrature settings and one color-basis setting provide the required correlations, with a number of settings independent of size.  Appendix~\ref{app:observable} specifies the normalization, mode-response requirement, and sampling assumptions.  Multicomponent alkaline-earth atoms offer another route through established $SU(N)$ exchange and correlation measurements~\cite{Gorshkov2010,CazalillaRey2014,Scazza2014,Taie2022}.  Ferromagnetic exchange or direct symmetric-state preparation is needed to realize the ground-state protocol there; the shell identity itself holds for either exchange sign.

\section{Conclusion and outlook}
\label{sec:outlook}

We have established a mechanism by which a quantum many-body ground state can heat faster than a continuous family of finite-temperature preparations.  Ferromagnetic exchange fixes the ground state to a color-Dicke wavefunction, whose configuration probabilities already equal those of the final mixed state.  The exact coherence-shell algebra prevents the evolution from generating population imbalances and fixes the ground-state decay rate.  For thermal states, a cumulative energy-band condition guarantees a slower contribution at every positive temperature.  Its exact finite-chain realization proves the separation across the full temperature range, and larger-chain calculations verify the criterion for both integrable and nonintegrable exchange dynamics.  The Gibbs family can therefore be continuous in state space while its zero-temperature endpoint has an isolated faster asymptotic rate.

Extending the cumulative sign condition to arbitrary size would determine how broadly this thermal ordering holds in the thermodynamic limit.  At fixed dephasing, that question also requires the size dependence of the thermal slow-mode amplitude: a small rate need not produce an observable crossing on accessible times.  Unequal color dephasing and exchange anisotropy require a corresponding calculation of the induced slow weight and the surviving acceleration interval.  Energy and color-correlation measurements in qutrit simulators provide finite-system tests of these predictions.  The exact ground-state subspace gives a starting point for controlling many-body heating through the relaxation modes accessed by a preparation.

\begin{acknowledgments}
This work was supported by the National Natural Science Foundation of China (Grant Nos.~12304290 and 12505017) and the Beijing National Laboratory for Condensed Matter Physics (Grant No.~2025BNLCMPKF017).  L.P. acknowledges support from the Fundamental Research Funds for the Central Universities (Grant No.~010-63263115).
\end{acknowledgments}

\appendix
\section{Asymptotic decay rate at finite size}
\label{app:proof}

For the initial deviation $X_0=\rho_0-\rhoss$, the finite-dimensional Jordan expansion on the traceless sector is
\begin{equation}
 \label{eq:jordan-expansion}
 e^{\Lio t}X_0=\sum_{\lambda\ne0}e^{\lambda t}
 \sum_{r=0}^{s_\lambda-1}\frac{t^r}{r!}N_\lambda^r\Pi_\lambda X_0.
\end{equation}
In Eq.~\eqref{eq:jordan-expansion}, $N_\lambda$ is nilpotent within its generalized eigenspace.  If $a$ is the largest real part among eigenvalues for which $\Pi_\lambda X_0\ne0$, the triangle inequality gives an upper envelope proportional to $(1+t^{s_{\max}-1})e^{at}$.  Projection onto the direct sum of eigenspaces with real part $a$ supplies the matching exponential lower envelope.  Polynomial Jordan factors and bounded oscillatory interference therefore leave the logarithmic limsup unchanged, so $-\limsup_{t\to\infty}t^{-1}\ln\|e^{\Lio t}X_0\|=-a=\GapF$.  At fixed finite size, norm equivalence makes the exponent independent of the chosen norm.  A thermodynamic limit requires separate control of the size-dependent prefactors and the order of limits.

\section{Analytic comb coefficients and spectral projections}
\label{app:counting}

The shell eigenvalue fixes a decay rate but does not fix the normalization of its eigenoperator.  We therefore distinguish the coefficient multiplying $O_m$ in Eq.~\eqref{eq:exactdynamics} from the coefficient in an orthonormal mode expansion.  Write $\nu_m(\bm N)$ for the number of configurations at distance $m$ from any one reference configuration in $\Omega_{\bm N}$.  Site permutations act transitively on this set, so this number is independent of the reference and the total ordered-pair count is $n_m=\mathcal D_{\bm N}\nu_m$.  Orthogonality of matrix units gives
\begin{equation}
 \label{eq:shell-orthogonality}
 \begin{aligned}
 \Tr(O_m^\dagger O_n)&=\delta_{mn}n_m,\\
 \|X_D\|_2^2&=1-\mathcal D_{\bm N}^{-1}.
 \end{aligned}
\end{equation}
For every occupied shell, including $m=0$, Eq.~\eqref{eq:shell-orthogonality} thus defines mutually orthonormal eigenoperators,
\begin{equation}
 \begin{aligned}
 R_m&=O_m/\sqrt{n_m},\qquad R_0=I_{\bm N}/\sqrt{\mathcal D_{\bm N}},\\
 \Lio R_m&=\Lio^\dagger R_m=-m\gamma R_m.
 \end{aligned}
 \label{eq:normalized-shell-modes}
\end{equation}
The simultaneous right and left eigenoperator property in Eq.~\eqref{eq:normalized-shell-modes} makes the shell modes their own Hilbert--Schmidt duals.  The ground-state expansion therefore has the exact amplitudes
\begin{equation}
 \begin{aligned}
 \rho_D(t)&=\sum_{m:\nu_m>0}c_m e^{-m\gamma t}R_m,\\
 c_m&=\Tr(R_m^\dagger\rho_D)
      =\frac{\sqrt{n_m}}{\mathcal D_{\bm N}}
      =\sqrt{\frac{\nu_m}{\mathcal D_{\bm N}}}.
 \end{aligned}
 \label{eq:analytic-comb-amplitudes}
\end{equation}
Equation~\eqref{eq:analytic-comb-amplitudes} includes the stationary coefficient $c_0=\mathcal D_{\bm N}^{-1/2}$.  For the transient $X_D$, this term is removed.  The two plotted weight conventions are consequently
\begin{equation}
 \begin{aligned}
 w_m&=|c_m|^2=\frac{\nu_m}{\mathcal D_{\bm N}},
       &\sum_{m\ge0}w_m&=1,\\
 p_{-m\gamma}&=\frac{|c_m|^2}{\|X_D\|_2^2}
       =\frac{\nu_m}{\mathcal D_{\bm N}-1},
       &\sum_{m\ge2}p_{-m\gamma}&=1.
 \end{aligned}
 \label{eq:comb-weight-normalizations}
\end{equation}
The first line of Eq.~\eqref{eq:comb-weight-normalizations} is used in Fig.~\ref{fig:comb}(a), while the second is Eq.~\eqref{eq:dicke-projection} and is used in Fig.~\ref{fig:modeprojections}(c).  By contrast, every nonzero $O_m$ has the same unnormalized expansion coefficient $1/\mathcal D_{\bm N}$ in Eq.~\eqref{eq:exactdynamics}.  Unequal comb weights arise from unequal shell norms.  The shell span is invariant under both $\Lio$ and $\Lio^\dagger$, so its orthogonal complement is invariant as well.  At a degenerate eigenvalue, $R_m$ selects the occupied direction within that eigenspace; other basis choices can redistribute individual mode coefficients.  The full spectral projection and its weight remain unchanged.  No orthogonality assumption for generic modes of the full non-normal generator is required.

It remains to evaluate $\nu_m$ without diagonalizing the Liouvillian.  For a fixed reference configuration with $N_a$ sites of color $a$, let $n_{ab}$ count positions that change from $a$ to $b$.  These nonnegative integers obey $\sum_b n_{ab}=N_a$, $\sum_a n_{ab}=N_b$, and $\sum_a n_{aa}=L-m$.  Choosing the target colors independently within each reference-color group gives
\begin{equation}
 \label{eq:shell-counting}
 \nu_m(\bm N)=\sum_{\{n_{ab}\}}\prod_a\frac{N_a!}{\prod_bn_{ab}!}.
\end{equation}
An equivalent generating polynomial makes the occupation constraint explicit:
\begin{equation}
 \begin{aligned}
 G_{\bm N}(z)&=\sum_{m=0}^L\nu_m z^m\\
 &=\left[\prod_a x_a^{N_a}\right]
   \prod_a\left(x_a+z\sum_{b\ne a}x_b\right)^{N_a}.
 \end{aligned}
 \label{eq:shell-generating-polynomial}
\end{equation}
Here square brackets denote extraction of the indicated monomial coefficient, rather than multiplication.  In Eq.~\eqref{eq:shell-generating-polynomial}, an unchanged site contributes $x_a$ and a changed site contributes $zx_b$.  Extracting the prescribed powers enforces the final occupations.  Thus $G_{\bm N}(1)=\mathcal D_{\bm N}$ and $G_{\bm N}(0)=1$, proving the normalizations in Eq.~\eqref{eq:comb-weight-normalizations}.  Equations~\eqref{eq:shell-counting} and~\eqref{eq:shell-generating-polynomial} are finite analytic expressions for arbitrary $SU(N)$ occupations, independent of the exchange graph and couplings.

For $SU(3)$, the matrix sum can be reduced to a short explicit sum.  Let $r_a=N_a-n_{aa}$ be the number of changed sites of reference color $a$, and set $t=n_{12}$.  Occupation conservation fixes all remaining off-diagonal entries:
\begin{equation}
 \begin{aligned}
 (n_{12},n_{13})&=(t,r_1-t),\\
 (n_{21},n_{23})&=(r_1+r_2-r_3-t,r_3-r_1+t),\\
 (n_{31},n_{32})&=(r_3-r_2+t,r_2-t).
 \end{aligned}
 \label{eq:su3-shell-flows}
\end{equation}
Choosing the $r_a$ changed positions and distributing their target colors using Eq.~\eqref{eq:su3-shell-flows} yields
\begin{equation}
 \begin{aligned}
 \nu_m(\bm N)&=\sum_{\substack{r_1+r_2+r_3=m\\0\le r_a\le N_a}}
     \left[\prod_{a=1}^3\binom{N_a}{r_a}\right] B(r_1,r_2,r_3),\\
 B(r_1,r_2,r_3)&=\sum_{t=0}^{r_1}\binom{r_1}{t}
                  \binom{r_2}{r_1+r_2-r_3-t}\\
 &\hspace{2em}\times\binom{r_3}{r_3-r_2+t}.
 \end{aligned}
 \label{eq:su3-explicit-shell-count}
\end{equation}
We use $\binom{u}{v}=0$ unless $0\le v\le u$.  This convention removes inadmissible flows without additional summation bounds.  Equation~\eqref{eq:su3-explicit-shell-count}, together with Eq.~\eqref{eq:analytic-comb-amplitudes}, gives the amplitude on every occupied $SU(3)$ comb tooth directly from $(N_1,N_2,N_3)$.  The first shells have a simple interpretation valid for arbitrary $N$:
\begin{equation}
 \begin{aligned}
 \nu_0&=1,\qquad \nu_1=0,\\
 \nu_2&=\sum_{a<b}N_aN_b,\\
 \nu_3&=2\sum_{a<b<c}N_aN_bN_c.
 \end{aligned}
 \label{eq:lowest-shell-counts}
\end{equation}
The two-site exchanges and the two orientations of a three-color cycle give the last two lines of Eq.~\eqref{eq:lowest-shell-counts}.  Hence the slowest occupied transient tooth is always $-2\gamma$ when at least two colors are present, and a third occupied color creates the odd tooth $-3\gamma$.  With only two colors, conservation requires equal numbers of both types of mismatch, recovering
\begin{equation}
 \nu_{2q}=\binom{N_1}{q}\binom{N_2}{q},\qquad \nu_{2q+1}=0.
 \label{eq:two-color-comb-counts}
\end{equation}
Equation~\eqref{eq:two-color-comb-counts} gives the XXZ Dicke-comb weights after the normalization in Eq.~\eqref{eq:comb-weight-normalizations}.  As a three-color example, $(N_1,N_2,N_3)=(2,2,2)$ gives $\mathcal D_{\bm N}=90$ and $(\nu_0,\ldots,\nu_6)=(1,0,12,16,27,24,10)$.  The squared amplitudes are these integers divided by $90$, while the transient weights are $(12,16,27,24,10)/89$.  This distinction also explains the two plotting conventions without invoking fitted mode amplitudes.

The $L=18$ coefficients in Fig.~\ref{fig:comb}(a) are evaluated by integer dynamic programming over the row allocations $n_{ab}$.  The count satisfies $\sum_m\nu_m=\mathcal D_{\bm N}$ exactly and reproduces the directly enumerated $L=6$ weights.  This evaluates the exact shell formula at a Hilbert-space dimension $\mathcal D_{(6,6,6)}=17153136$; it does not require the full Liouvillian spectrum.  Panel (b) shows direct trace-norm calculations through $L=9$.

For an independent five-site calculation, we form the complete $900\times900$ Liouvillian for $L=5$ and $\bm N=(2,2,1)$.  If $V$ contains its right eigenvectors, the rows of $V^{-1}$ are their biorthogonal left duals.  Repeated eigenvalues are grouped before the projection is evaluated,
\begin{equation}
 \label{eq:biorthogonal-projector}
 \Pi_\lambda=V E_\lambda V^{-1},\qquad
 (E_\lambda)_{\alpha\alpha}=
 \begin{cases}1,&\lambda_\alpha=\lambda,\\0,&\lambda_\alpha\ne\lambda\end{cases}.
\end{equation}
The projector in Eq.~\eqref{eq:biorthogonal-projector} is invariant under basis changes within a degenerate eigenspace.  We use an absolute eigenvalue-grouping tolerance $2\times10^{-8}$ in units of $J_1$; it is well below the minimum separation of the distinct spectral groups in either calculation.  For the two couplings, reconstruction of $X_D$ from the complete projectors has relative error below $1.3\times10^{-13}$, and the largest absolute difference from the analytic comb weights is below $3.1\times10^{-15}$.  The largest off-comb squared projection is below $1.7\times10^{-25}$ and is consistent with roundoff.  The raw eigenvalues, grouped projections, shell counts, and validation summary accompany the reconstruction script.

For the seven-site spectra in Fig.~\ref{fig:modeprojections}, we resolve every character of translation and the occupation-preserving cyclic color permutation.  All character blocks enter the calculation, and their dimensions sum to the full $\mathcal D_{\bm N}^2$.  At $\bm N=(3,2,2)$ there are fourteen blocks, each of dimension $3150$.  Direct diagonalization of each block therefore yields the complete spectrum at both exchange choices without a large dense inverse eigenvector matrix.  We evaluate the projection of $X_D$ within its exact reducing shell span.  If $\mathcal M=\{m\ge2:O_m\ne0\}$, then
\begin{equation}
 \Pi_{-m\gamma}X_D=
 \prod_{\substack{n\in\mathcal M\\n\ne m}}
 \frac{\Lio+n\gamma}{(n-m)\gamma}\,X_D.
 \label{eq:reducing-polynomial-projection}
\end{equation}
Equation~\eqref{eq:reducing-polynomial-projection} is restricted to the shell reducing space containing $X_D$.  Projection of a general operator in the full non-normal space would require the full spectral projectors.  The resulting weights agree with the shell counts.  We check the block representation against the full generator at four, five, and six sites.  The complete six-site dense spectrum provides a further benchmark.

\section{Spectral formulation and mode selection}
\label{app:selection}

The long-time decay of $X_0=\rho_0-\rhoss$ is set by the slowest spectral subspace on which $X_0$ has a nonzero projection.  In a fixed strong-symmetry sector with a unique steady state, it is convenient to denote this visible rate by
\begin{equation}
 \GapF(\rho_0)=
 -\max_{\lambda\ne0:\,\Pi_\lambda X_0\ne0}\Re\lambda .
 \label{eq:filteredgap}
\end{equation}
In Eq.~\eqref{eq:filteredgap}, $\Pi_\lambda$ is the Riesz projector onto the generalized eigenspace at $\lambda$.  For a diagonalizable generator, this is the usual selection by left-eigenoperator overlaps; the projector form also covers degeneracies without choosing a basis inside the slow space~\cite{Prosen2008,Minganti2018}.  This notation records the usual dependence of the observed rate on the initial-state projection.  For a distance induced by a norm, two states display an asymptotic strong Mpemba inversion when
\begin{equation}
 \label{eq:mpemba-condition}
 \begin{aligned}
 D(\rho_A,\rhoss)&>D(\rho_B,\rhoss),\\
 \GapF(\rho_A)&>\GapF(\rho_B),
 \end{aligned}
\end{equation}
since their distance ratio then vanishes exponentially.  Appendix~\ref{app:proof} shows that Jordan blocks leave this logarithmic rate unchanged.  The exact shell algebra in Sec.~\ref{sec:colorcomb} determines the excluded projections directly from the generator.

Symmetry provides one part of this construction.  If $[\Lio,\mathcal U_g]=0$ and the initial density matrix is invariant under the group action, all left modes in nontrivial operator representations are absent~\cite{BucaProsen2012,AlbertJiang2014,SanchezMunoz2019,BucaTindall2019}.  Scalar slow modes may then be removed by matching their expectation values to those of the stationary state.  If $\{Q_\mu\}$ spans the remaining slow left subspace, the required condition is
\begin{equation}
 \label{eq:moment-matching}
 \Tr\!\left[Q_\mu^\dagger(\rho_A-\rhoss)\right]=0
 \quad\text{for every }\mu .
\end{equation}
Suppose the relevant observables close under the adjoint dynamics,
\begin{equation}
 \label{eq:adjoint-closure}
 \Lio^\dagger(Q_\mu)=\sum_\nu M_{\nu\mu}Q_\nu .
\end{equation}
Equations~\eqref{eq:moment-matching} and~\eqref{eq:adjoint-closure} reduce the cancellation check to a finite operator algebra.  The XXZ comb and the exact Hamiltonian mode of the constrained chain illustrate these two forms of selection.  Symmetry-based acceleration also occurs in dissipative condensate preparation~\cite{Westhoff2025}, while conserved quantities and typicality provide other ways to restrict the accessible dynamics~\cite{Ulcakar2025,LiuWang2025,Bao2026}.  A rate inversion requires excluding every mode slower than the leading reference contribution.  Any remaining slow projection would set the final exponent.

This emphasis on asymptotic state-distance decay fixes the scope of the comparison with other QME diagnostics.  Non-normal evolution can produce transient crossings without a rate inversion~\cite{LonghiNonnormal2025}, and relaxation-trajectory lengths define a different ordering from Eq.~\eqref{eq:mpemba-condition}~\cite{QianIntrinsic2025}.  Operator-based Mpemba diagnostics have also been developed for adjoint evolution~\cite{BaguiOperators2026}.  Spectral control has also been explored in dissipative incommensurate and mosaic systems~\cite{XuIncommensurate2025,DongMosaic2025,ZhouQuasiperiodic2026}.  Resource-theoretical descriptions relate thermal and symmetry-restoration effects through mode visibility~\cite{SummerResource2025}, while coherent preparation, measurement-based steering, and temporary reset protocols can suppress slow components~\cite{BeatoControl2026,HongGeometric2026,BaoReset2025}.  Related many-body effects after bosonic quenches~\cite{YamashikaBosons2025}, in imaginary-time dynamics~\cite{ChangImaginary2026}, and for circuit quantum resources~\cite{AdityaComplexity2026} need not share a Lindblad state-distance exponent; the same qualification applies to temperature crossings in strongly coupled chaotic models~\cite{WangChaotic2026} and to descriptions using Ruelle--Pollicott resonances in closed systems~\cite{YamashikaResonances2026}.  The present comparison uses one fixed Markovian generator.  Its exact eigenoperators and the thermal sign criterion establish the rate separation.

\section{Proof and verification for every finite positive temperature}
\label{app:alltemperature}

The criterion in Sec.~\ref{sec:thermal} uses complete energy projectors and is therefore independent of the eigenvector basis within a degeneracy.  Since $Q$ is a left eigenoperator at a nonzero eigenvalue, stationarity of $\rhoss$ gives $\Tr Q=0$.  Equation~\eqref{eq:suncomb} also gives $\Tr(Q\rho_D)=0$ whenever $\Gamma_s$ differs from every occupied comb rate.  Thus $a_0=A_M=0$.  Applying finite summation by parts to the Gibbs expansion yields Eq.~\eqref{eq:abel-positive}; nonnegative $A_k$, with one strict inequality, make it positive for every $\beta>0$.  The conclusion requires neither a uniform positive lower bound as $T\to0,\infty$ nor an exchange of the long-time and temperature limits.  The thermal overlap behaves as
\begin{equation}
 \begin{aligned}
 F_Q(\beta)&=-\frac{\beta}{\mathcal D_{\bm N}}\Tr(QH)+O(\beta^2),
                   &&\beta\to0,\\
 F_Q(\beta)&\sim a_*e^{-\beta(E_*-E_0)},&&\beta\to\infty,
 \end{aligned}
 \label{eq:thermal-overlap-endpoints}
\end{equation}
where $E_*$ is the first excited energy with nonzero $a_k$.  The high-temperature coefficient in Eq.~\eqref{eq:thermal-overlap-endpoints} has a fixed sign because $\Tr(QH)=-\sum_{k=1}^{M-1}A_k(E_{k+1}-E_k)<0$.  The low-temperature expression also explains why the thermal tail can become exponentially weak without disappearing at a fixed positive temperature.  Its competition with a fast term gives a tail-dominance scale proportional to $\beta(E_*-E_0)/(2\gamma-\Gamma_s)$; this need not equal the first Mpemba crossing between two nearby preparations.

Trace-norm duality gives Eq.~\eqref{eq:thermal-distance-bound}.  Shell orthogonality also gives $D_{\rm tr}[\rho_D(t),\rhoss]\le\sqrt{\mathcal D_{\bm N}-1}\,e^{-2\gamma t}/2$.  Combining the two bounds makes the asymptotic inversion explicit,
\begin{equation}
 \frac{D_{\rm tr}[\rho_D(t),\rhoss]}{D_{\rm tr}[\rho_\beta(t),\rhoss]}
 \le\frac{\|Q\|_\infty\sqrt{\mathcal D_{\bm N}-1}}{F_Q(\beta)}
 e^{-(2\gamma-\Gamma_s)t}.
 \label{eq:explicit-ratio-bound}
\end{equation}
The prefactor in Eq.~\eqref{eq:explicit-ratio-bound} is finite for each fixed $\beta>0$.  This proves eventual overtaking without assuming diagonalizability of the remaining Liouvillian.  Moreover, the Hermitian part of the generator yields
\begin{equation}
 \frac{\|\operatorname{diag}Q\|_2^2}{\|Q\|_2^2}
 \ge1-\frac{\Gamma_s}{2\gamma}>0.
 \label{eq:slow-population-content}
\end{equation}
Inequality~\eqref{eq:slow-population-content} follows because every nonzero off-diagonal matrix unit in the fixed-color sector has Hamming distance at least two.  Thus a sub-comb witness necessarily carries population weight, quantifying the channel separation used in the physical interpretation.

For the real exchange matrices used here, the column-vectorized Liouvillian is complex symmetric, $\Lio^{\mathsf T}=\Lio$.  A Hermitian left mode $Q$ at a simple real eigenvalue has the right partner $Q^*$, with nonzero biorthogonal denominator $\Tr(QQ^*)$.  Its contribution to the energy is
\begin{equation}
 \delta E_s(t)=
 \frac{\Tr(HQ^*)F_Q(\beta)}{\Tr(QQ^*)}\,e^{-\Gamma_s t}.
 \label{eq:energy-witness-response}
\end{equation}
The numerator in Eq.~\eqref{eq:energy-witness-response} is nonzero: since $H$ is real symmetric, $\Tr(HQ^*)=\Tr(QH)\ne0$ under the sign condition.  The same selected mode therefore appears in energy and in the full state.  The simple-eigenvalue assumption is needed for this scalar energy formula; state-distance visibility from a nonzero left-mode overlap does not require it.  Numerically we check the energy overlap, bilinear normalization, and eigenoperator residual in addition to the cumulative signs.

\subsection{An exact finite-chain example}
\label{app:thermal-exact}

Take $L=4$, $\bm N=(2,1,1)$, $J_2=0$, and $\gamma/J_1=2/5$.  The fixed-color Hilbert space has dimension twelve.  Simultaneous translation, reflection, and interchange of the equally populated colors reduce its invariant operator space to thirteen orbit sums.  In unnormalized orbit-mass coordinates, the restricted generator has rational real and imaginary entries.  Its characteristic polynomial has a simple real root $z_s=\lambda_s/J_1$ in $(-0.584,-0.583)$, belonging to
\begin{equation}
 \begin{aligned}
 p(z)={}&78125z^8+625000z^7+6175000z^6\\
       &+28660000z^5+109967500z^4\\
       &+272501200z^3+354979040z^2\\
       &+219987456z+50595840.
 \end{aligned}
 \label{eq:exact-witness-polynomial}
\end{equation}
The root of Eq.~\eqref{eq:exact-witness-polynomial} therefore gives $0<\Gamma_s/J_1=-z_s<4/5=2\gamma/J_1$.  The remaining factors of the thirteen-dimensional characteristic polynomial are $z(z+4/5)^2(z+6/5)(z+8/5)$.  An exact Routh count for $p(z-3/5)$ has one first-column sign change, so exactly one root of $p$ lies to the right of $\Re z=-3/5$.  It is the isolated root $z_s$ above.  All other nonstationary roots are therefore faster, establishing the leading exponent in the common invariant sector.  The energy levels in units of $J_1$ are $(-4,-2,0,2)$, with multiplicities $(1,5,3,3)$.  The residue of the energy-resolved resolvent at $z_s$, followed by a fixed normalization of $Q$, gives
\begin{equation}
 (a_0,a_1,a_2,a_3)=(0,1,R-1,-R).
 \label{eq:exact-energy-band-coefficients}
\end{equation}
The ratio $R$ in Eq.~\eqref{eq:exact-energy-band-coefficients} is determined algebraically by
\begin{equation}
 \begin{aligned}
 A(z)={}&3125z^5+20000z^4+100500z^3\\
       &+322800z^2+448400z+210816,\\
 R={}&1+\frac{62500z_s(z_s+24/25)(z_s+7/5)}{A(z_s)}.
 \end{aligned}
 \label{eq:exact-ratio-expression}
\end{equation}
Sturm root counting isolates one root in the rational interval $(-583176616733,-583176616732)/10^{12}$.  Substitution in Eq.~\eqref{eq:exact-ratio-expression} gives $A(z_s)>0$ and $7/10<R<4/5$, so $(A_1,A_2)=(1,R)$ is strictly positive.  This establishes Eq.~\eqref{eq:exact-all-temperature-overlap} for every $\beta>0$, with $Z_\beta=e^{4\beta J_1}+5e^{2\beta J_1}+3+3e^{-2\beta J_1}$.  Its energy response is nonzero because $\Tr(QH)=-2J_1(1+R)$.  The exact-arithmetic program constructs the orbit generator and all energy projectors from the bonds, derives the resolvent residues, and verifies the inequalities directly.

\subsection{Larger-chain sign checks}
\label{app:thermal-numerics}

For the size extension we build the orbit-normalized Liouvillian in the nearest-balanced sectors, resolving translation, reflection, and occupation-preserving color permutations.  The quotient dimensions are $13$, $50$, $147$, $1620$, $9986$, $26418$, and $442160$ for $L=4$--$10$.  Dense diagonalization is used for the smaller quotients and shift-invert eigenvectors at $L=8$.  At $L=9,10$, we instead apply largest-real-part Arnoldi iteration after deflating the exact shell span, avoiding a complete sparse LU factorization.  Each Hermitian witness has $\|Q\|_2=1$ and is checked against the original adjoint generator.  The full sector Hamiltonian is diagonalized, degenerate energies are combined, and every energy-band trace is evaluated.  Table~\ref{tab:alltemperature} gives the selected rates and minimum cumulative margins.  Their implication for every $\beta>0$ follows from Eq.~\eqref{eq:abel-positive}.  The larger-chain results are finite-precision sign checks; the four-site calculation supplies the exact certificate.

Figure~\ref{fig:heating} uses the $26418$-dimensional quotient at $L=9$.  Each finite-temperature Gibbs deviation is propagated directly, and the energy is contracted in the orbit basis.  The trace distance is evaluated from all translation blocks of the evolved deviation, an exact unitary block decomposition.  Initial trace distances agree with the Boltzmann-weight expression.  Ground-state curves use Eq.~\eqref{eq:exactdynamics}, whose selected trace distances are independently checked by full-matrix diagonalization.  The extended curves, checkpoints, and time-step comparison accompany the source; the long-time precision control is described in Appendix~\ref{app:numerics}.

\begin{table}[!tbp]
\caption{\label{tab:alltemperature}Energy-band sign checks at $\gamma/J_1=0.4$.  Witnesses have $\|Q\|_2=1$, and $A_{\rm min}$ is the minimum over $1\le k<M$.  The four-site nearest-neighbor sign is also established independently by exact arithmetic.  The $J_2/J_1=0.6$ cases at $L=4,5$ contain vanishing cumulative entries within roundoff and are omitted from this strict-interior-margin sequence; their raw tables are retained.}
\begin{ruledtabular}
\begin{tabular}{ccccc}
$L$ & $J_2/J_1$ & $\Gamma_s/J_1$ & $A_{\rm min}$ & $\Tr(QH)/J_1$\\
\hline
4 & 0 & 0.583177 & 0.75583 & $-3.5876$\\
5 & 0 & 0.571181 & 0.49050 & $-8.7048$\\
6 & 0 & 0.560207 & 0.18875 & $-17.8490$\\
6 & 0.6 & 0.520494 & 0.23424 & $-12.6267$\\
7 & 0 & 0.563401 & 0.20601 & $-30.1162$\\
7 & 0.6 & 0.562520 & 0.21909 & $-26.3688$\\
8 & 0 & 0.553757 & 0.12538 & $-53.5190$\\
8 & 0.6 & 0.560215 & 0.14093 & $-51.1324$\\
9 & 0 & 0.548455 & 0.05773 & $-99.3971$\\
9 & 0.6 & 0.557477 & 0.06740 & $-100.1427$\\
10 & 0 & 0.541167 & 0.04588 & $-162.8807$\\
10 & 0.6 & 0.554445 & 0.05233 & $-171.6508$\\
\end{tabular}
\end{ruledtabular}
\end{table}

All listed eigenoperator residuals are below $5\times10^{-14}$, and the smallest positive cumulative margin remains above $0.04$ in this normalization.  At $L=9,10$, the independent MATLAB/Octave and Python implementations agree on the full energy-band cumulative curves.  For $L=10$, a separate direct application of the full-space adjoint to the stored witness also verifies its residual, independently of the quotient assembly.  Energy grouping is repeated at one quarter and four times the default tolerance before a point is accepted.  Repeated energy eigenvalues are combined before the traces are taken; no sign is assigned to a basis-dependent member of a degenerate energy space.  Raw coefficients, cumulative sums, witness matrices or orbit vectors, and numerical residuals accompany the scripts.  Eigenstate data for smaller sizes and native large-size checkpoints are stored separately.  The selected rates are eigenvalue bounds in Eq.~\eqref{eq:all-temperature-rate}.  For example, the $L=8$ next-nearest-neighbor witness rate differs from the production finite-window tail fit by about $7\times10^{-5}$ relatively.  This difference measures a subleading-tail bias, separately from the propagation and fitting-window errors, and leaves the strict separation from $2\gamma$ intact.

\subsection{A graph on which the thermal contrast vanishes}
\label{app:thermal-complete-graph}

The exact ground-state comb alone cannot imply a rate inversion on every exchange graph.  For uniform complete exchange, $H=-J\sum_{i<j}P_{ij}$ is invariant under every site permutation, as is every Gibbs state.  The permutation action is transitive on the fixed-color configurations, so the Gibbs diagonal is uniform.  Each of its Hamming components also commutes with every exchange.  Its evolution is therefore confined to dephasing shells just as for the Dicke state.  For $\beta>0$, strict positivity of the matrix exponential of the connected ferromagnetic exchange graph gives nonzero distance-two entries, and both preparations have leading rate $2\gamma$.  The shell identity is therefore graph independent, while the existence of a slower thermal contribution depends on the graph.  Both the rate and sign conditions are needed for the continuous-temperature comparison.

\section{Numerical implementation and validation}
\label{app:numerics}

The fixed-color basis is enumerated without duplicate permutations, and each exchange bond is applied to every configuration to construct the sparse Hamiltonian.  At the largest sizes, propagation uses the matrix-free Liouvillian action.  The energy trajectories in Fig.~\ref{fig:heating} use the exact symmetry quotient described in Appendix~\ref{app:alltemperature}; the complete spectral check in Fig.~\ref{fig:modeprojections} uses all translation and color-character blocks, as described in Appendix~\ref{app:counting}.  With column-major vectorization the full action is
\begin{equation}
 \label{eq:liouvillian-vectorization}
 \begin{aligned}
 \Lio|\rho\rangle\rangle=-i(I\otimes H-H^{\mathsf T}\otimes I)
 |\rho\rangle\rangle\\
 {}-\gamma\,\mathrm{diag}(G)|\rho\rangle\rangle.
 \end{aligned}
\end{equation}
Here $G_{xy}=d_H(x,y)$; matrix-free propagation of Eq.~\eqref{eq:liouvillian-vectorization} uses an Arnoldi--Krylov exponential~\cite{Saad1992,Sidje1998}.  For $L=6$, $\bm N=(2,2,2)$, and $J_2=0.6$, the largest normalized shell-commutator residual is below $5.2\times10^{-16}$, and the Frobenius difference between full Liouvillian evolution and Eq.~\eqref{eq:exactdynamics} is $1.3\times10^{-16}$ at a representative time.  The exact-sparse validation data are listed in Table~\ref{tab:data}, while the additional mode-projection calculation is documented in Appendix~\ref{app:counting}.

\begin{table}[tb]
\caption{\label{tab:data}Exact-sparse trajectory data for $\gamma=0.4$ and $\beta=0.25$.  Rates are finite-window exponential estimates, not exact eigenvalues.  The first crossing time, in units of $J_1^{-1}$, is interpolated between adjacent grid points.  Exact modal witnesses and their distinction from fitted tails are given in Appendix~\ref{app:alltemperature}.}
\begin{ruledtabular}
\begin{tabular}{ccccccc}
$L$ & $\bm N$ & $\mathcal D$ & $J_2$ & $\Gamma_T$ & $\eta$ & $t_\times$\\
\hline
4 & $(2,1,1)$ & 12  & 0    & 0.5875 & 1.361 & 10.66\\
5 & $(2,2,1)$ & 30  & 0    & 0.5714 & 1.400 & 8.62\\
6 & $(2,2,2)$ & 90  & 0    & 0.5601 & 1.428 & 8.53\\
6 & $(2,2,2)$ & 90  & 0.60 & 0.5199 & 1.539 & 9.14\\
7 & $(3,2,2)$ & 210 & 0.60 & 0.5624 & 1.422 & 9.55\\
\end{tabular}
\end{ruledtabular}
\end{table}

For the matrix-free sequence, the initial state is $\rho_{\rm HT}=[I-a_L(H-\bar H I)]/\mathcal D$, where $\bar H=\Tr(H)/\mathcal D$ and $a_L\|H-\bar H I\|_\infty\le0.8$.  It is positive and has the same first-order direction as $e^{-\beta H}/Z$ at $\beta=0$, while the normalized tail is independent of $a_L$.  The normalized Hilbert--Schmidt tail contains $37$ time points.  The default fit uses $10^{-8}<\|\rho-\rhoss\|_2/\|\rho(0)-\rhoss\|_2<10^{-3}$ and is repeated over the five windows $(10^{-7},10^{-3})$, $(10^{-8},10^{-3})$, $(10^{-8},10^{-4})$, $(10^{-9},10^{-3})$, and $(3\times10^{-8},3\times10^{-4})$.  Table~\ref{tab:matrixfree} reports the large-size rates and, for each $L$, the larger fit-window variation and in-window Krylov residual of the two coupling choices.  Fits to the exact $\beta=0.25$ Gibbs trace-distance curves at $L=8,9$ differ from the corresponding tangent-state rates by at most $1.3\times10^{-4}$ relatively; the four comparisons are included in the archived data.

\begin{table*}[t]
\caption{\label{tab:matrixfree}Matrix-free finite-$\gamma$ rate audit for the high-temperature Gibbs tangent at $\gamma=0.4$.  Here $\eta_{\rm HT}=2\gamma/\Gamma_{\rm HT}$, $\delta_{\rm win}$ is the maximum relative change of $\Gamma_{\rm HT}$ across the five fit windows, and $R_{\rm K}$ is the maximum relative Krylov residual inside the default fit window.}
\begin{ruledtabular}
\begin{tabular}{cccccccc}
$L$ & $\mathcal D$ & $\Gamma_{\rm HT}(0)$ & $\eta_{\rm HT}(0)$ & $\Gamma_{\rm HT}(0.6)$ & $\eta_{\rm HT}(0.6)$ & $\max\delta_{\rm win}$ & $\max R_{\rm K}$\\
\hline
8  & 560   & 0.553766 & 1.44465 & 0.560256 & 1.42792 & $5.6\times10^{-5}$ & $4.8\times10^{-11}$\\
9  & 1680  & 0.548458 & 1.45864 & 0.557500 & 1.43498 & $3.3\times10^{-5}$ & $7.8\times10^{-11}$\\
10 & 4200  & 0.541168 & 1.47828 & 0.554462 & 1.44284 & $2.7\times10^{-5}$ & $1.9\times10^{-10}$\\
11 & 11550 & 0.530516 & 1.50796 & 0.547793 & 1.46040 & $1.9\times10^{-5}$ & $1.1\times10^{-9}$\\
\end{tabular}
\end{ruledtabular}
\end{table*}

For a direct convergence test, we repeated the nonintegrable $L=8$, $\bm N=(3,3,2)$, $J_2/J_1=0.6$ point for Krylov dimensions $m=20$, $30$, and $40$, and repeated $m=30$ and $40$ after halving every propagation interval.  Relative to the $m=40$ half-step reference, the production choice $m=30$ changes neither the fitted curve nor the rate above double-precision roundoff within the same implementation.  More conservatively, an independent reconstruction agrees with the stored production curve within $2.4\times10^{-7}$ pointwise in the fit window and within $1.5\times10^{-8}$ in the fitted rate.  The comparison table and reproduction script accompany the numerical data.

The extended Gibbs trajectories in Figs.~\ref{fig:heating} and~\ref{fig:dynamics} are obtained by sparse exponential actions in the invariant quotient, through $\gamma t=20$.  We propagate the traceless deviation $X=\rho-\rhoss$ and renormalize it after each interval while retaining its accumulated scale.  The exact stationary component is projected out at roundoff level, preventing stationary contamination from setting the apparent tail.  Trace norms include all momentum blocks of the translation-invariant density matrix; direct full-matrix diagonalizations at selected times provide an independent check of this block evaluation.  The Dicke trace norm is calculated from Eq.~\eqref{eq:exactdynamics} with $e^{-2\gamma t}$ factored out before diagonalization.  No thermal tail is extrapolated from an exponential fit.  A repeated $L=9$, $J_2/J_1=0.6$ calculation with half the propagation interval gives the same extended curve within accumulated roundoff; the pointwise comparison and checkpoints are archived with the data.

The temperature scans in Fig.~\ref{fig:windows}(a,b), together with the archived $L=5$ data, confirm the rate inversion over $0.005\le\beta\le2$, while the crossing time is controlled by the slow-mode amplitude.  Exact trace-distance trajectories were evaluated independently in the nearest-balanced $L=7$--$9$ sectors at the nine temperatures listed in the data tables.  The main tail amplitudes use the common window $10^{-9}<D_{\rm tr}<10^{-3}$.  At $L=9$, repeating the fit over $10^{-8}<D_{\rm tr}<10^{-4}$ and $10^{-9}<D_{\rm tr}<10^{-5}$ changes the high-temperature amplitude exponent from $0.9869$ to $0.9888$ and $0.9893$, respectively.  The late-window thermal rates over $\beta\le0.1$ lie between $0.548454$ and $0.548462$.  Crossings are linearly interpolated between adjacent points on a grid with $\Delta(\gamma t)=0.1$; the grid spacing, rather than the printed decimal expansion, sets their numerical resolution.  The dashed lines test the expected linear-amplitude and logarithmic-delay forms rather than exact coefficients.

The checkpointed $L=8$ and $L=9$ density matrices preserve unit trace to $2\times10^{-15}$ and Hermiticity to $2.1\times10^{-16}$ relatively.  Internal Arnoldi error estimators that fall below double-precision roundoff are not interpreted as total numerical accuracy; only conservative convergence tolerances and directly checkable invariants are used in the claims above.

\section{Population relaxation in the strong-dephasing limit}
\label{app:zenogap}

The strong-dephasing limit gives an effective description of slow population relaxation in the nearest-neighbor chain.  For $\gamma\gg J$, coherences decay before the populations change appreciably.  Eliminating them to second order in $J/\gamma$ yields a classical exchange generator.  This quantum Zeno limit is a complementary analysis; the ground-state shell identity and the finite-$\gamma$ thermal-overlap criterion hold without this approximation.  Let $\mathcal P$ project onto the color-basis diagonal, $\mathcal Q=1-\mathcal P$, and $\Lio_H=-i[H,\cdot]$.  Adiabatic elimination gives
\begin{equation}
 \label{eq:zeno-elimination}
 \Lio_{\rm eff}=-\mathcal P\Lio_H\mathcal Q
 (\mathcal Q\Diss\mathcal Q)^{-1}\mathcal Q\Lio_H\mathcal P.
\end{equation}
In Eq.~\eqref{eq:zeno-elimination}, a nearest-neighbor exchange creates a distance-two coherence decaying at $2\gamma$; the two conjugate virtual paths therefore produce
\begin{equation}
 \label{eq:zeno-classical-generator}
 K=w\sum_j(P_{j,j+1}-I),\qquad w=J^2/\gamma.
\end{equation}
Equation~\eqref{eq:zeno-classical-generator} uses the convention $\partial_t p=Kp$.  We restrict to the nearest-balanced three-color sequence $|N_a-N_b|\le1$ and project onto translations, reflection, and the global color permutations that preserve the occupation multiset.  A one-density-wave excitation is absent from this scalar quotient, so the first candidate contains opposite momenta.  In the relative coordinate $r=1,\ldots,L-1$, the nonconstant component obeys a discrete Laplacian with reflecting endpoints.  The total-momentum-zero Bethe condition is $e^{ik(L-1)}=1$~\cite{Sutherland1975}; after excluding the stationary $k=0$ multiplet, reflection and color scalarity select
\begin{equation}
 \label{eq:zeno-relative-branch}
 \begin{aligned}
 f_r&\propto\cos\!\left[\frac{2\pi}{L-1}\left(r-\frac12\right)\right],\\
 \lambda_1&=-4w\left[1-\cos\!\left(\frac{2\pi}{L-1}\right)\right].
 \end{aligned}
\end{equation}
The branch in Eq.~\eqref{eq:zeno-relative-branch} has decay rate
\begin{equation}
 \Gamma_T^{\rm inv}=\frac{4J^2}{\gamma}
 \left[1-\cos\!\left(\frac{2\pi}{L-1}\right)\right]
 \sim\frac{8\pi^2J^2}{\gamma(L-1)^2}.
 \label{eq:giantscaling}
\end{equation}
Equation~\eqref{eq:giantscaling} refers to the effective population generator, after taking the strong-dephasing limit; it is not an exact finite-$\gamma$ rate.  We compare this branch with the lowest nonstationary rates of the orbit-normalized scalar quotient.  For $L=5$--$18$ the candidate is the restricted gap within numerical precision, with maximum absolute discrepancy $5.4\times10^{-15}$.  Establishing that it remains the lowest branch at arbitrary $L$ would additionally require excluding lower Bethe-string branches.

At high temperature, the diagonal deviation is linear in $\beta$, with bias given by the diagonal part of $H-\langle H\rangle_0$ up to quadratic corrections.  We define its normalized slow weight as the Euclidean norm of the projection onto the complete first scalar eigenspace divided by the norm of the projected bias.  It is nonzero for every $L=5$--$18$ but decreases from $1$ to $0.2577$ [Fig.~\ref{fig:thermochaos}(b)].  The projection is basis independent within the degenerate eigenspace.  Its decreasing weight leaves the thermodynamic-limit amplitude undetermined.

\section{Fully symmetry-resolved level statistics}
\label{app:levelstats}

Schur--Weyl duality decomposes $(\mathbb C^3)^{\otimes L}$ into $SU(3)$ irreducible representations labeled by Young diagrams with at most three rows.  In the multiplicity space of a fixed diagram $\lambda$, adjacent transpositions act in the Young orthogonal representation as
\begin{equation}
 \label{eq:young-orthogonal}
 s_i|T\rangle=d_i^{-1}|T\rangle+
 \sqrt{1-d_i^{-2}}|s_iT\rangle.
\end{equation}
In Eq.~\eqref{eq:young-orthogonal}, $d_i$ is the content difference between boxes containing $i+1$ and $i$, and the second term is omitted when the swapped tableau is nonstandard.  Products of these sparse matrices represent every exchange $P_{ij}$ and the lattice translation $\mathcal T$.

We select momentum with the standard group projector built from powers of $\mathcal T$, orthonormalize its range, and project the Hamiltonian.  Complex conjugation maps $k$ to $-k$, as does reflection, so the antiunitary product $\Theta=\mathcal I\mathcal K$ acts within each selected block and obeys $\Theta^2=1$.  The appropriate Wigner--Dyson comparison is therefore GOE.  We then remove the lowest and highest $12\%$ of each block and form adjacent-gap ratios without unfolding or mixing blocks~\cite{Atas2013}.  The two Young shapes and positive momenta used at each size are $(6,4,2),(5,4,3)$ with $k=1,\ldots,5$ for $L=12$; $(6,4,3),(7,4,2)$ with $k=1,\ldots,6$ for $L=13$; $(7,4,3),(6,5,3)$ with $k=1,\ldots,6$ for $L=14$; and $(7,5,3),(8,4,3)$ with $k=1,\ldots,7$ for $L=15$.  The retained ratio counts are $1500$, $4350$, $10089$, and $28388$ per coupling.  Uncertainties use the corresponding $10$, $12$, $12$, and $14$ block means as independent samples.  Across $L=12$--$15$, the integrable means remain within $0.3851$--$0.3887$, whereas the $J_2/J_1=0.6$ means lie within $0.5294$--$0.5432$.  The former agree with Poisson and the latter are GOE-like, although the largest-size value remains slightly below $4-2\sqrt3$.  Removing $8\%$ or $16\%$ of each spectral edge at $L=12$ leaves this classification unchanged.

\section{Observable visibility}
\label{app:observable}

For a measured observable $A$, the deviation is $\delta A(t)=\sum_\alpha c_\alpha a_\alpha e^{\lambda_\alpha t}$ with response amplitude $a_\alpha=\Tr(AR_\alpha)$.  An observable detects the filtered-gap hierarchy only if it responds to the leading visible modes of both preparations.  State-level trace distance avoids this additional filter, but scalable experiments and tensor-network calculations generally do not.  To compare both preparations using the same readout, center the population observable and normalize both contributions by their operator norms,
\begin{equation}
 \label{eq:normalized-readout}
 \begin{aligned}
 E_{{\rm cl},0}&=E_{\rm cl}-\Tr(E_{\rm cl})I_{\bm N}/\mathcal D_{\bm N},\\
 \widetilde A_\kappa&=\frac{C_2}{\|C_2\|_\infty}
                    +\kappa\frac{E_{{\rm cl},0}}{\|E_{{\rm cl},0}\|_\infty}.
 \end{aligned}
\end{equation}
The traces and norms in Eq.~\eqref{eq:normalized-readout} are restricted to the chosen fixed-color sector.  The parameter $\kappa$ is fixed before comparing the two preparations; both signal magnitudes are evaluated as $|\Tr[\widetilde A_\kappa(\rho(t)-\rhoss)]|$.  This observable has a response to the exact coherence shell and the population tail, but its crossing is a separate measurable consequence rather than a replacement for the state-distance test.

To obtain the digital dephasing channel, take the phases in $U_\phi$ to be independent, zero-mean real Gaussian variables with variance $\gamma\delta t$.  The average for two color configurations is
\begin{equation}
 \label{eq:averaged-phase-channel}
 \overline{\langle x|U_\phi\rho U_\phi^\dagger|y\rangle}
 =\exp[-\gamma\delta t\,d_H(x,y)]\rho_{xy}.
\end{equation}
Expanding Eq.~\eqref{eq:averaged-phase-channel} to first order gives Eq.~\eqref{eq:dephasing}.  A common phase for all colors at a site is physically irrelevant, while a spatially uniform color-dependent field produces collective noise; neither substitutes for independent local relative phases.  Replacing the isotropic covariance by a positive semidefinite color covariance matrix generates unequal color-pair rates.  For $N=3$, $C_2$ is measured in the six bases associated with $X^{01}$, $Y^{01}$, $X^{02}$, $Y^{02}$, $X^{12}$, and $Y^{12}$, while $E_{\rm cl}$ is measured directly in a seventh, color-basis setting.  All pairs $i<j$ entering $C_2$ can be accumulated from the same site-resolved record, so the number of settings does not grow with $L$.  With independent shots, each bounded-correlator uncertainty scales as $M^{-1/2}$; correlations among pair estimators from one shot must still be retained in the uncertainty of their sum.

For the archived readout example, exact sparse Liouvillian evolution is performed in the balanced $L=9$ sector at $J_2/J_1=0.6$, $\gamma/J_1=0.4$, $\beta J_1=0.25$, and $\kappa=2$.  The same fixed observable is evaluated for both preparations.  The source archive includes these trajectories and additional readout scans.  They illustrate observable sensitivity; the state-level SQME follows from the spectral criterion and trace-distance calculations.

\bibliographystyle{apsrev4-2}
\bibliography{refs}

\end{document}